\documentclass[aps,prx,preprint,superscriptaddress]{revtex4-2}

\usepackage{graphicx}
\usepackage{dcolumn}
\usepackage{bm}
\usepackage[font=small,labelfont=bf]{caption}
\usepackage{float}
\usepackage{tabularx}
 
\begin{document}

\title{Time-resolved chiral X-Ray photoelectron spectroscopy with transiently enhanced atomic site-selectivity: a Free Electron Laser investigation of electronically excited fenchone enantiomers}

\author{D.~Faccialà}
\email{Corresponding authors: davide.facciala@polimi.it, michele.devetta@cnr.it, caterina.vozzi@cnr.it}
\affiliation{Istituto di Fotonica e Nanotecnologie - CNR (CNR-IFN), Milano, Italy}

\author{M.~Devetta}
\email{Corresponding authors: davide.facciala@polimi.it, michele.devetta@cnr.it, caterina.vozzi@cnr.it}
\affiliation{Istituto di Fotonica e Nanotecnologie - CNR (CNR-IFN), Milano, Italy}

\author{S.~Beauvarlet}
\affiliation{Université de Bordeaux-CNRS-CEA, CELIA, Talence, France}

\author{N.~Besley}
\email{Deceased}
\affiliation{School of Chemistry\char`,{} University of Nottingham\char`,{} Nottingham, United Kingdom}

\author{F.~Calegari}
\affiliation{Center for Free-Electron Laser Science\char`,{} DESY\char`,{} Hamburg, Germany}
\affiliation{Hamburg University\char`,{} Physics Department\char`,{} Hamburg, Germany}

\author{C.~Callegari}
\affiliation{Elettra-Sincrotrone Trieste, Basovizza, Italy}

\author{D.~Catone}
\affiliation{Istituto di Struttura della Materia - CNR (ISM-CNR), Roma, Italy}

\author{E.~Cinquanta}
\affiliation{Istituto di Fotonica e Nanotecnologie - CNR (CNR-IFN), Milano, Italy}

\author{A.~G.~Ciriolo}
\affiliation{Istituto di Fotonica e Nanotecnologie - CNR (CNR-IFN), Milano, Italy}

\author{L.~Colaizzi}
\affiliation{Center for Free-Electron Laser Science\char`,{} DESY\char`,{} Hamburg, Germany}

\author{M.~Coreno}
\affiliation{Elettra-Sincrotrone Trieste, Basovizza, Italy}
\affiliation{Istituto di Struttura della Materia - CNR (ISM-CNR), Roma, Italy}

\author{G.~Crippa}
\affiliation{Istituto di Fotonica e Nanotecnologie - CNR (CNR-IFN), Milano, Italy}
\affiliation{Dipartimento di Fisica\char`,{} Politecnico di Milano\char`,{} Milano, Italy}

\author{G.~De~Ninno}
\affiliation{Elettra-Sincrotrone Trieste, Basovizza, Italy}
\affiliation{Laboratory of Quantum Optics\char`,{} University of Nova Gorica, Slovenia}

\author{M.~Di~Fraia}
\affiliation{Elettra-Sincrotrone Trieste, Basovizza, Italy}

\author{M.~Galli}
\affiliation{Center for Free-Electron Laser Science\char`,{} DESY\char`,{} Hamburg, Germany}
\affiliation{Dipartimento di Fisica\char`,{} Politecnico di Milano\char`,{} Milano, Italy}

\author{G.~A.~Garcia}
\affiliation{Synchrotron Soleil\char`,{} Gif sur Yvette, France}

\author{Y.~Mairesse}
\affiliation{Université de Bordeaux-CNRS-CEA, CELIA, Talence, France}

\author{M.~Negro}
\affiliation{Istituto di Fotonica e Nanotecnologie - CNR (CNR-IFN), Milano, Italy}

\author{O.~Plekan}
\affiliation{Elettra-Sincrotrone Trieste, Basovizza, Italy}

\author{P.~Prasannan~Geetha}
\affiliation{Istituto di Fotonica e Nanotecnologie - CNR (CNR-IFN), Milano, Italy}
\affiliation{Dipartimento di Fisica\char`,{} Politecnico di Milano\char`,{} Milano, Italy}

\author{K.~C.~Prince}
\affiliation{Elettra-Sincrotrone Trieste, Basovizza, Italy}

\author{A.~Pusala}
\affiliation{Istituto di Fotonica e Nanotecnologie - CNR (CNR-IFN), Milano, Italy}
\affiliation{Dipartimento di Fisica\char`,{} Politecnico di Milano\char`,{} Milano, Italy}

\author{S.~Stagira}
\affiliation{Istituto di Fotonica e Nanotecnologie - CNR (CNR-IFN), Milano, Italy}
\affiliation{Dipartimento di Fisica\char`,{} Politecnico di Milano\char`,{} Milano, Italy}

\author{S.~Turchini}
\affiliation{Istituto di Struttura della Materia - CNR (ISM-CNR), Roma, Italy}

\author{K.~Ueda}
\affiliation{Institute of Multidisciplinary Research for Advanced Materials, Tohoku University, Sendai, Japan}

\author{D.~You}
\affiliation{Institute of Multidisciplinary Research for Advanced Materials, Tohoku University, Sendai, Japan}

\author{N.~Zema}
\affiliation{Istituto di Struttura della Materia - CNR (ISM-CNR), Roma, Italy}

\author{V.~Blanchet}
\affiliation{Université de Bordeaux-CNRS-CEA, CELIA, Talence, France}

\author{L.~Nahon}
\affiliation{Synchrotron Soleil\char`,{} Gif sur Yvette, France}

\author{I.~Powis}
\affiliation{School of Chemistry\char`,{} University of Nottingham\char`,{} Nottingham, United Kingdom}

\author{C.~Vozzi}
\email{Corresponding authors: davide.facciala@polimi.it, michele.devetta@cnr.it, caterina.vozzi@cnr.it}
\affiliation{Istituto di Fotonica e Nanotecnologie - CNR (CNR-IFN), Milano, Italy}

\begin{abstract}
Chiral molecules are widespread in nature, playing a fundamental role in bio-chemical processes and in the origin of life itself. The observation of dynamics in chiral molecules is crucial for the understanding and control of the chiral activity of photo-excited states. One of the most promising techniques for the study of photo-excited chiral systems is time-resolved photoelectron circular dichroism (TR-PECD), which offers an intense and sensitive probe for vibronic and geometric molecular structure as well as electronic structures, and their evolution on a femtosecond timescale. However, the non-local character of the PECD effect, which is imprinted during the electron scattering off the molecule, makes the interpretation of TR-PECD experiments challenging. In this respect, core-photoionization is known to allow site- and chemical-sensitivity to photelectron spectroscopy. Here we demonstrate that TR-PECD utilising core-level photoemission enables probing the chiral electronic structure and its relaxation dynamics with atomic site sensitivity. Following UV pumped excitation to a 3s Rydberg state, fenchone enantiomers (C$_{10}$H$_{16}$O) were probed on a femtosecond scale using circularly polarized soft X-ray light pulses provided by the free-electron laser FERMI. C 1s binding energy shifts caused by the redistribution of valence electron density in this 3s-valence-Rydberg excitation allowed us to measure transient PECD chiral responses with an enhanced C-atom site-selectivity compared to that achievable in the ground state molecule.
These results represent the first chemical-specific and site-specific, enantio-sensitive observations on the electronic structure of a photo-excited chiral molecule and pave the way towards chiral femtochemistry probed by core-level photoemission.
\end{abstract}
\maketitle

\section{\label{sec:introduction} Introduction}
The study of chirality is of paramount importance across the physical and biological sciences. Chirality, or handedness, is the property of an object that breaks mirror symmetry, making it non-superimposable to its mirror image. The two mirror images of the same chiral compound are called enantiomers, or optical isomers. These can be distinguished by their interaction with another chiral entity, such as circularly polarized light or chiral reagents. In spite of having the same chemical structure, enantiomers may then show a class of distinctive chemical properties, via the so-called chiral recognition fundamental process. Of continuing interest, terrestrial life is homochiral at the molecular level \cite{Bonner95,meierhenrich2008} with, for instance,
all the naturally occurring chiral amino acids in proteins found to present the same left-handed configuration.  Consequently, the biological activity and perception of the many chiral pharmaceuticals \cite{h2011significance} and odorants \cite{Brenna03} can be very enantiomer dependent.

A chiroptical effect that constitutes one of the most promising investigation techniques for studying chiral compounds in the gas phase is photoelectron circular dichroism (PECD) \cite{powis2008photoelectron}. In PECD, circularly polarized radiation ionizes a randomly oriented enantiomeric sample, revealing its chirality by a strong asymmetry between the intensity of electron emission in the forward and backward direction along the propagation axis of the ionizing radiation beam. This asymmetry reverses direction with exchange of the enantiomer or with the handedness of the circularly polarized light (CPL). The dichroism, or normalized difference between the left- and right-handed CPL angular distributions, is used to define a PECD chiral asymmetry factor. As opposed to conventional circular dichroism, which relies on weak magnetic dipole and electric quadrupole processes, PECD arises in a pure electric dipole approximation and consequently produces much stronger chiral asymmetries of up to a few tens percent. This makes PECD ideally suited for the study of dilute matter \cite{powis2008photoelectron,turchini2017review,hadidi2018electron}. PECD can also be effectively exploited for probing the molecular dynamics of a chiral sample with enantiosensitivity, which is very important for the understanding and manipulation of chiral interactions. In this case, the molecule is excited with a short laser pulse (pump) and the subsequent temporal dynamics is probed with a second photoionizing probe pulse. Recently the first UV time-resolved PECD (TR-PECD) valence-shell studies in chiral molecules have been obtained \cite{comby2016relaxation, beaulieu2016probing, blanchet2021ultrafast}, demonstrating the sensitivity of this technique as a probe of the ultrafast intra-molecular relaxation taking place in a chiral sample excited in a Rydberg state. Indeed, TR-PECD could provide access to information on dynamics related to alignment, vibrational relaxation and electronic conversion.

In these pioneering PECD experiments, the observed asymmetries from the photoelectron emission might be thought to arise from either the chirality of its initial orbital or the chirality of the surrounding effective potential through which it scatters. Disentangling the role of these two contributions can be experimentally challenging.
One way to address this issue is to single-photon ionize the molecule directly from a highly localized atomic core level with an X-ray or soft X-ray probe. The initial orbital in this case is chemically specific, potentially localized and, for an \emph{s orbital}, spherically symmetric, that is achiral. It thus serves as an ideal in-situ probe of the influence of photoelectron scattering by the chiral molecular potential. Furthermore, depending on the molecular structure, different localization sites may be distinguishable by their binding energies, due to different chemical shifts, and as such simultaneously probed with a single experimental measurement. Compared to valence PECD, core-level PECD thus provides a powerful and direct enantio-sensitive probe of the electronic scattering potential of a molecule that is both site- and chemical- specific, and thus more sensitive to the electronic and structural environment surrounding the initial localized core state. The application of core-level PECD spectroscopy to the study of chirality of molecules in their ground state has proven to be a very sensitive tool to investigate the structure of molecules \cite{hergenhahn2004, harding2005, ulrich2008giant,powis2008glycidol,turchini2013conformational,ilchen2017emitter,tia2017,ilchen2021site, nalin2021photoelectron, fehre2021fourfold}.
However, time-resolved core-level PECD studies clearly require the use of ultrashort, circularly polarized X-ray probe pulses, and such developments have been restricted by the limited availability of suitable high-brilliance sources in the soft X-ray range. In fact, only one time-resolved measurement, a chemical-specific investigation of a dissociating chiral cation (C$_3$H$_3$F$_3$O$^+$), has been reported \cite{ilchen2021site} so far. Here, the measurement of PECD at a single binding energy corresponding to the fluorine K-edge was impeded because the contribution of the three F-atom sites in this molecule could not be disentangled and so the overall observed PECD presented a limited average value. Due to the very challenging nature of this experiment, an unambiguous PECD could only be observed for a single time delay \cite{ilchen2021site}.
Recently, Mayer et al. \cite{mayer2022following} have demonstrated that time-resolved X-ray Photoelectron Spectroscopy (TR-XPS) of electronically excited states provides access to the charge distribution and ultrafast relaxation pathways in a chemical-selective way. By monitoring the transient excited-state chemical shift (ESCS) in the S 2p$^{-1}$ core electron binding energy in photo-excited 2-thiouracil, the transient local charge density around the sulphur S atom was probed. Transient ESCS triggered by photo-excitation could potentially enhance the site-specificity of TR-XPS techniques, especially in the case where several atoms of the same chemical element are probed simultaneously. This capability is particularly appealing for studying organic compounds at the Carbon K-edge, where it could be exploited for distinguishing the contribution of different carbon atoms inside the molecule through their different ESCS.

In the present work, we combine the chemical- and site- specificity of TR-XPS with the enantio-specificity of TR-PECD, for studying the photo-excited Rydberg state of fenchone, a prototypical organic compound, at the carbon K-edge. We performed the experiment at the FERMI free electron laser (FEL) light source, the only FEL providing circularly polarized soft X-ray pulses up to the C K-edge with high temporal coherence. The FEL probe pulse was used in combination with a jitter-free external laser used as a UV pump.
With this proof-of-principle experiment, we show how the electronic excitation provided by the UV pump not only initiates molecular dynamical processes in the electronically excited molecule , but also causes  transient ESCS in the core electron binding energies that originate from the redistribution of the valence electron density after photo-excitation. In particular, we demonstrate how these transient shifts permit enhancing the site specificity of the technique, by allowing to further separate and identify individual, and subsets of, C-atom sites in the C 1s$^{-1}$ spectra that, otherwise, would not be accessible.
The results of our experiment are complemented with advanced theoretical calculations, allowing us to assign and partially decipher the contribution of the different carbon photoemission sites to the overall TR-PECD signal. 
This establishes a new, enantio-sensitive, site- and chemical- specific experimental tool for electron spectroscopy of photoexcited chiral species, that we call TR-chiral-XPS.

\section{\label{sec:experiment} Experiment}
\begin{figure*}[!ht]
	\centering
	\includegraphics[width = 1\textwidth]{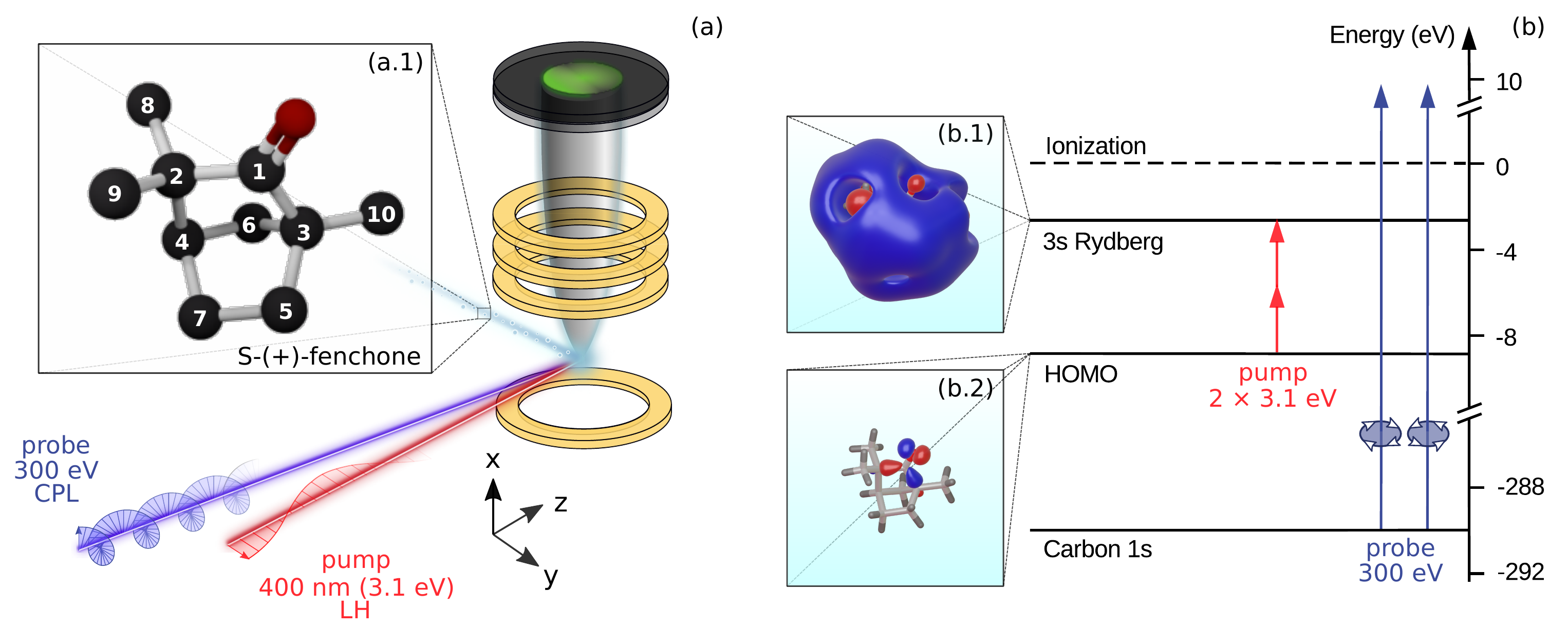}
	\caption{(a) Experimental setup. The circularly polarized XUV probe (CPL) and the visible pump with linear horizontal (LH) polarization are focused quasi-collinearly in the molecular beam of S-(+)-fenchone enantiomers. (a.1) 3D representation of the S-(+)-fenchone, where only carbon (black) and oxygen (red) atoms are shown, while hydrogen atoms are hidden for the sake of clarity. Carbon atoms are labeled from 1 to 10. (b) Simplified energy level diagram of the excitation scheme. (b.1) 3D representation of the LUMO+1 orbital (3s Rydberg state). (b.2) 3D representation of the HOMO orbital. Representations	(a.1), (b.1) and (b.2) share the same molecular orientation.}
	\label{fig:0}
\end{figure*}

Fenchone molecule (C$_{10}$H$_{16}$O; Fig. \ref{fig:0}a.1) was chosen as an easy-to-handle, prototypical rigid chiral system, that has been used as a benchmark in previous one-photon valence-shell \cite{powis2008valence, nahon2016determination}, core-shell \cite{ulrich2008giant}
and UV multi-photon
\cite{lux2015photoelectron, comby2016relaxation, beaulieu2016probing, kastner2017intermediate, blanchet2021ultrafast} PECD studies. Most importantly, fenchone exhibits a giant chiral asymmetry in the C 1s core-level photoemission from its ground electronic state \cite{ulrich2008giant}. In the carbon K-edge region, the X-ray photoelectron spectrum (XPS) shows two peaks separated by $\sim\,2.3\,$eV with that at higher binding energy identified as the carbonyl (C=O) carbon 1s$^{-1}$ ionization. This 2.3 eV chemical shift is rationalized as a de-screening at this C$_1$ atomic site caused by the electron withdrawing inductive effect of the adjacent electronegative O atom. All the remaining C atom sites in this molecule sit in very similar chemical environments to one another, leading to their having closely similar binding energies that are not resolved; they hence appear as a single XPS peak of 1$\,$eV bandwidth.

The outermost (HOMO) valence orbital of fenchone, nominally an oxygen lone pair, is in fact highly localized around the C$_2$-C$_1$(=O)-C$_3$ atom grouping (Fig. \ref{fig:0}b.2).
In our experiment, photoexcitation by a linearly polarized ultrafast visible pump pulse with photon energy E$_{\mathrm{pump}}$=3.1$\,$eV excited the localized HOMO into the diffuse 3s Rydberg state by two-photon absorption, triggering ultrafast relaxation dynamics \cite{comby2016relaxation}. It has been shown that a two-photon absorption greatly enhances the relative excitation cross-section to the 3s state \cite{singh2020}.
When an electron from this localized orbital is promoted into a diffuse molecular Rydberg state by UV pump-photoexcitation (Fig. \ref{fig:0}b.1), one may expect that the reduced valence electron density at the C$_1$, C$_2$ and C$_3$ atoms will cause similar chemical shifts to higher core-level binding energy for all three carbons. Hence, we anticipate up-shifted C$_{1-3}$ 1s$^{-1}$ XPS peaks, making the three carbon sites distinguishable from all others in core ionization from the excited 3s Rydberg state; as a corollary their presence provides a signature identifying the Rydberg-excited nature of the molecule being core-ionized.

The excited state and the subsequent relaxation dynamics were probed in the wake of excitation by circularly polarized FEL probe pulses with photon energy E$_{\mathrm{probe}}$=300$\,$eV. A scheme of the experimental setup, described in details in Sec. \ref{sec:methods:expset}, is shown in Fig.\ref{fig:0}a and a simplified energy level diagram representing the pump-probe excitation scheme is shown in Fig.\ref{fig:0}b. The energy and angular distribution of the photoelectrons emitted from the C K-edge were measured with a velocity map imaging (VMI) spectrometer generating 2D projections of the 3D photoelectron angular distribution on a position sensitive detector. Measurements were repeated, using alternating FEL helicities, at different pump-probe delays. 
The photoelectron angular distribution (PAD) and photoelectron spectrum (PES) were then retrieved from inverse-Abel transformed images as discussed in Sec. \ref{sec:methods:data1}. The radial coordinate maps the electron intensity onto the PES energy axis while the energy dependent PAD is expressed by fitting Legendre polynomials to the axially symmetric angular distribution at each radius. 

\section{\label{sec:staticPECD} Core-level PECD of Fenchone}
\begin{figure*}[!ht]
	\centering
	\includegraphics[width = 1\textwidth]{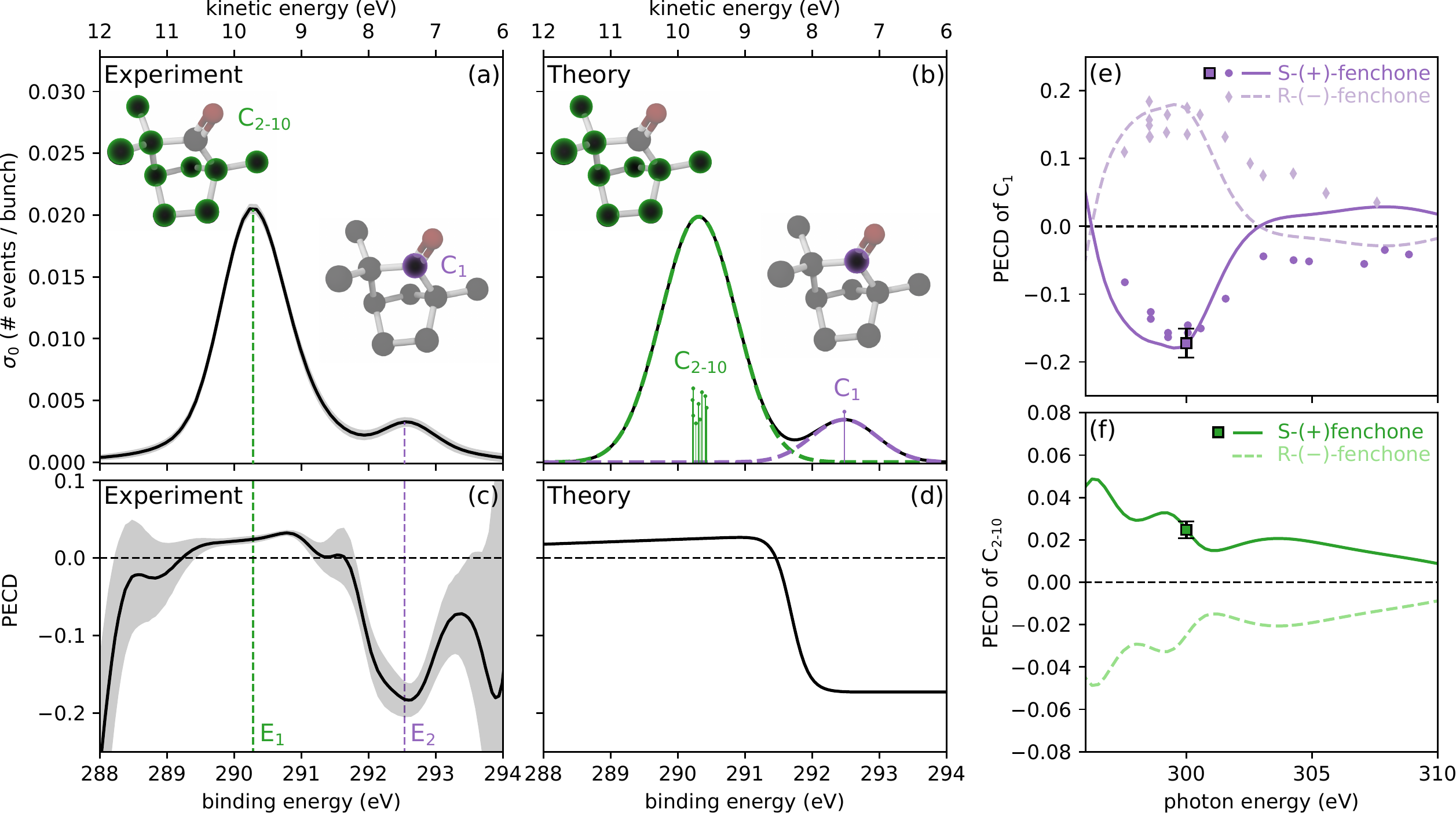}
	\caption{C K-edge ionization of ground state S-(+)-fenchone: (a) experimental PES recorded at 300 eV photon energy; and (b) a theoretically predicted spectrum based on the calculated C 1s binding energies. These are marked with green and violet sticks, the lengths of which are intended for clarity only. The gray area in (a) represents the $1\sigma$ confidence interval, multiplied by 10 for clarity. The associated experimental PECD results appear in (c), while in (d) the theoretically predicted PECD for S-(+)-fenchone is presented. The gray area in (c) represents the $1\sigma$ confidence interval. In both (c) and (d), the backward asymmetry from the C$_1$ localized emission reaches almost 20$\%$. 
	Theoretical PECD asymmetries for S-(+)-fenchone (solid line) and R-($-$)fenchone (dashed line) as a function of photon energy are presented for the C$_{1}$ peak in (e) and the unresolved average of the C$_{2-10}$ peaks is presented in (f). Dots and diamonds in (e) indicate experimental synchrotron data for S-(+)-fenchone and R-($-$)-fenchone, respectively (Ref. \cite{ulrich2008giant}). The experimental $h\nu=300$\ eV results from our current work for the  C$_{1}$ and C$_{2-10}$ peaks are shown with a violet and green square dots in (e) and (f), respectively, along with the corresponding $1\sigma$ error-bars.}
	\label{fig:1}
\end{figure*}

We first present in Fig.\ref{fig:1} the results obtained in the FEL probe-only \emph{ground state} ionization, where a gas-phase sample of randomly oriented S-(+)-fenchone enantiomers was probed with circularly polarized FEL pulses (degree of polarization $\ge 95\%$, see \cite{roussel2017polarization}) at 300$\,$eV. These results served as a benchmark for assessing the reliability and accuracy of core-level PECD measurements in our experiment.
The experimental PES $\sigma_0$ in Fig.\ref{fig:1}(a) shows  two distinct peaks as expected. The weaker one, indicated by a violet dashed line, is identified as the carbonyl C 1s peak (C$_1$ in Fig. \ref{fig:0}a), whose binding energy is shifted with respect to the stronger peak, indicated by a green dashed line, coming from all the remaining C 1s states (C$_{2-10}$ in Fig.\ref{fig:0}a). The experimental PES is simulated as a sum of gaussians centered at the theoretical ground state binding energies computed using a Density Functional Theory (DFT) $\Delta$SCF calculation with PBE0 functional and 6-311++G** basis. The relative intensity of the two apparent peaks, C$_{2-10}$ and C$_1$, has been adjusted to match the experimental PES;
this fit is shown in Fig. \ref{fig:1}(b), and the theoretical binding energies are indicated by vertical solid lines in Fig. \ref{fig:1}(b), with the violet and green lines indicating the C$_1$  and C$_{2-10}$ contributions, respectively.

Under these experimental conditions (single photon ionization, randomly oriented sample) the chiral antisymmetry of the PAD adopts a simple form, $b_1\cos \theta$, that can be fitted using the single adjustable parameter, $b_1$, and a PECD asymmetry factor (normalized difference between forward and backward photoelectron emission) can then be expressed as $2b_1$ \cite{powis2008photoelectron}. This experimentally measured PECD asymmetry is shown in Fig. \ref{fig:1}(c) as a function of binding energy and kinetic energy and agrees well with the theoretical PECD shown in Fig. \ref{fig:1}(d).
The experimental peaks observed at the binding energies $E_1 = 290.28\,$eV and $E_2 = 292.53\,$eV are indicated in Fig. \ref{fig:1}(c) by vertical dashed green and violet lines, respectively. The expected average PECD in correspondence of peak $E_j$, $\langle{\mathrm{PECD}}\rangle_j$, is estimated as the average of PECD$(E)$ over a window of energies of $\pm0.4\,$eV around the peak, weighting the average by the Gaussian one-electron response at $E_j$ (See Sec. \ref{sec:methods:data1} for further details). $\langle{\mathrm{PECD}}\rangle_2$ and $\langle{\mathrm{PECD}}\rangle_1$ are shown with squares in Figs. \ref{fig:1}(e) and  \ref{fig:1}(f), respectively, and compared with both the theoretical calculations and previous experimental results obtained in a synchrotron radiation study of fenchone\cite{ulrich2008giant}.

The comparison shows very good agreement both with the theory and previous experimental results, demonstrating that core-level spectroscopy allows accessing the chiral response of the molecular ground state probed from two different regions of the molecule, one constituted by the carbonyl C$_1$ peak and the other one represented by the weighed contribution from other carbons C$_{2-10}$. Note also that, both in the experimental and theoretical results, the PECD is shown to slightly increase across the C$_{2-10}$ peak when going to higher binding energies (Figs. \ref{fig:1}(c) and \ref{fig:1}(d)), which reflects the cumulative effect from all the C$_{2-10}$ contributions having different chiral asymmetries and binding energies.
The very good agreement with previous synchrotron-based data shows that high quality data can be acquired with the present FEL-based set-up despite its lower repetition rate.

\section{\label{sec:trpecd}  Time-resolved C $1s^{-1}$ ionization of $n \rightarrow 3s$\ excited fenchone}
We then adapted this approach for the time-resolved study of the relaxation dynamics of photo-excited fenchone. The molecules were excited with a linearly polarized visible pump pulse, with a full-width at half maximum (FWHM) duration $\sim\,75\,$fs, and were then probed by the FEL circularly polarized pulse, with a FWHM duration around $\sim\,15\,$fs, at seven different pump-probe delays in the temporal window ranging from $-200$ to $1000\,$fs. Then, the photoelectron angular distribution (PAD) and photoelectron spectrum (PES) for each pump-probe delay were retrieved from inverse-Abel transformed images as discussed in Sec. \ref{sec:methods:data}. The quasi-collinear geometry of these laser sources with a linear and a circular polarization breaks the axial symmetry at the sample, raising issues when processing the 2D VMI data using an Abel inversion technique, which presupposes this symmetry. While alternative detection methods, such as fast time and position sensitive detection \cite{vredenborg2008tofdetect,tia2017,nalin2021photoelectron} or tomographic reconstruction \cite{hockett2010tomo, lux2015photoelectron}
allow measuring the full 3D distribution without Abel inversion, for technical reasons their use was unfeasible in the present situation. However, as discussed in Sec. \ref{sec:methods:data2}, the use of the inverse Abel transform is completely justified in our case, since limited effects on the main outcomes are predicted. 

\subsection{\label{sec:sec:PES} Time-resolved PES}
\begin{figure*}[!ht]
	\centering
	\includegraphics[width = 0.9\textwidth]{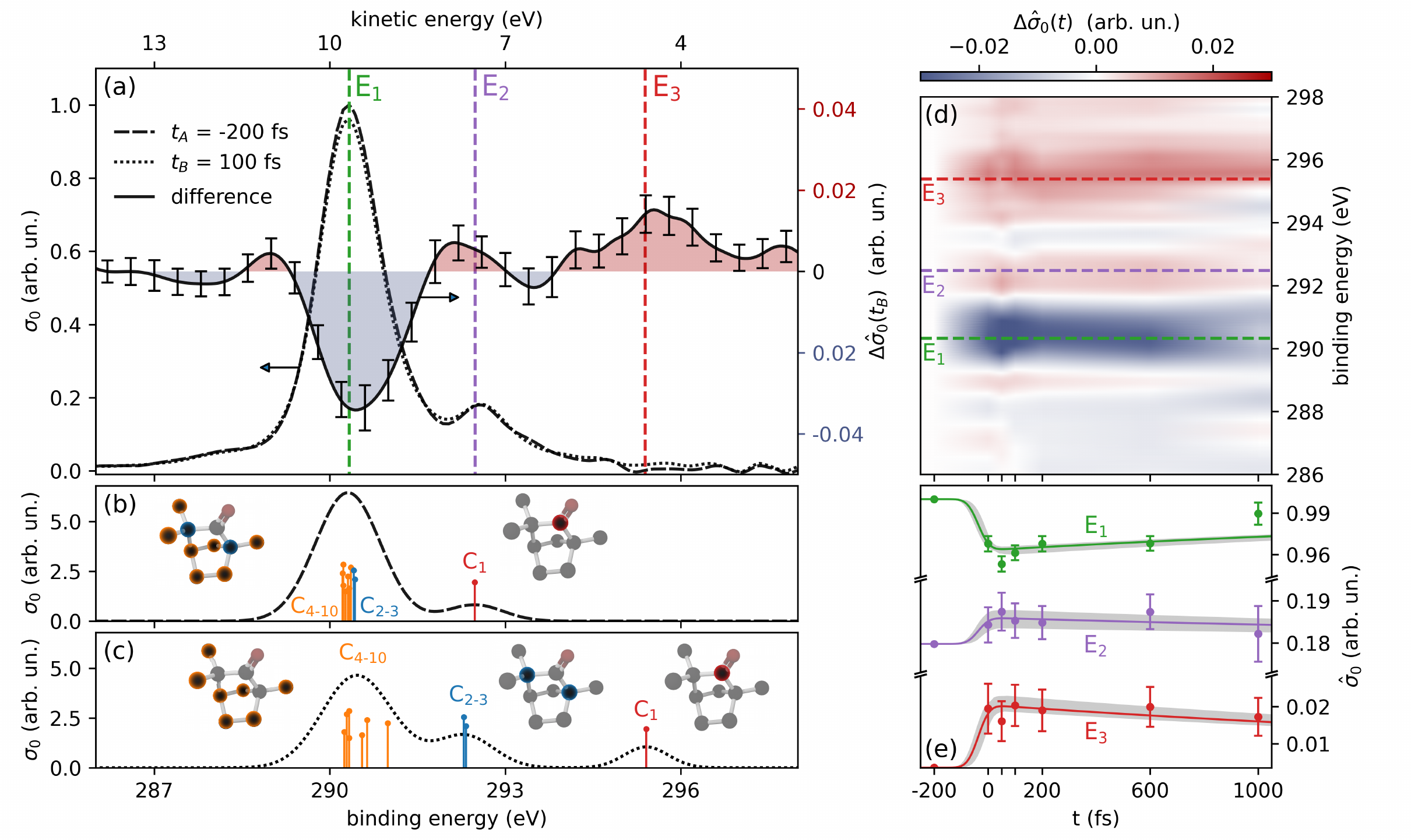}
	\caption{(a) PES at $t_A = -200\,$fs (dashed line), and $t_B=100\,$fs (dotted line), difference $\Delta \hat\sigma_0(t_B) = \hat\sigma_0(t_B) - \hat\sigma_0(t_A)$ of the sliding averaged PES $\hat\sigma_0$ (solid line and $1\sigma$ error-bars). (d) $\hat\Delta\sigma_0(t)$ for all the measured delays $t$. (a,d) The energies $E_1$, $E_2$ and $E_3$ are indicated by dashed green, violet and red lines, respectively. (e) $\hat\sigma_0(t)$ for $E_1$, $E_2$ and $E_3$ indicated by green, violet and red dots, respectively, along with the $1\sigma$ error-bars. A single exponential of 3.3 ps decay time, corresponding to the measured lifetime of the 3s Rydberg state \cite{comby2016relaxation} is shown with solid lines. The amplitude and time zero of this exponential are fitted to the experiment and the $1\sigma$ confidence interval of the fit is represented by a gray shaded area. (b,c) Calculated binding energies for the ground state (b) and excited 3s Rydberg state (c), indicated by red (C$_1$), blue (C$_{2}$, C$_3$) and orange (from C$_{4}$ to C$_{10}$) dots and vertical lines, where the vertical offset is intended for clarity only. Theoretical PES in (b) and (c) are shown with dashed and dotted line, respectively.  }
	\label{fig:2}
\end{figure*}

Figure \ref{fig:2}(a) compares the measured PES $\sigma_0$, normalized by its integral, at the pump-probe delays $t_A=-200\,$fs and $t_B = 100\,$fs, represented by a dashed and dotted line, respectively.
A sliding average of $1\,$eV is applied to the PES, corresponding to the energy resolution of the post-processed VMI images (1.2 eV at 8 eV kinetic energy, see Sec \ref{sec:methods:fitting}). We then evaluated the difference between the sliding averaged spectra $\hat\sigma_0$ corresponding to the two time delays: $\Delta \hat\sigma_0(t_B)= \hat\sigma_0(t_B) - \hat\sigma_0(t_A)$. This is shown with a solid line in Fig. \ref{fig:2}(a). The measurement reveals a clear depletion of the main peak at $E_1$. A small enhancement of the PES in the region of the peak at $E_2$, and a higher enhancement of the signal around $E_3 = 295.4\,$eV are also observed. 
This behavior is confirmed for all delays $t\ge0$: the general trend of $\Delta \hat\sigma_0(t)$ with time is shown in Fig. \ref{fig:2}(d) as a false color plot.
Three single-pixel lineouts at $E_1$, $E_2$, and $E_3$ of the sliding averaged spectra $\hat\sigma_0(t)$ are indicated in Fig. \ref{fig:2}(e) with green, purple and red dots, respectively. The amplitude of these variations is rather small, due to the low number of molecules excited in the 3s state by the two photon UV transition.
In our experimental conditions, we estimated an excitation of the fenchone molecules in the 3s Rydberg state of around $12.5\%$ as it is discussed in Sec. \ref{sec:methods:fitting}.
In Fig. \ref{fig:2}(e) the lineouts are compared with single exponential of 3.3 ps decay time. This decay time is consistent with the depletion of population of the 3s Rydberg state for $E_3$. For $E_1$, this time decay corresponds to a regrowth of the ground state population, with a measured lifetime of 3.3 ps  \cite{comby2016relaxation}.

These observations accord well with the initially stated expectations that the changes in valence electron density following the $n \rightarrow 3s$ Rydberg excitation would cause additional C 1s binding energy shifts at the C$_1$, C$_2$, and C$_3$ atomic sites that are adjacent to the localised valence HOMO orbital. 
Theoretically predicted C 1s binding energies in the ground and excited 3s Rydberg state are indicated in Figs. \ref{fig:2}(b) and \ref{fig:2}(c), respectively, and after folding with a shaping function provide theoretical PES that are overlaid in these panels as a black line. We notice that when the molecule is excited in the 3s Rydberg state, the carbonyl C$_1$\ 1s$^{-1}$ peak, shown in red in Figs. \ref{fig:2}(b,c), shifts to a higher binding energy of 295.41$\,$eV. 
Following a similar trend, the C$_2$ and C$_3$ carbon peaks, shown in blue in Figs. \ref{fig:2}(b,c), shift to around 292.3$\,$eV, being unfortunately overlapped with the binding energy of the C$_1$ peak for an emission taking place from the ground state. All the other contributions, shown in orange in Figs. \ref{fig:2}(b,c), slightly spread in energy but still within our experimental energy resolution. Thus, the depletion observed at $E_1$ in Figs. \ref{fig:2}(a,d-e) can be attributed to the shift of the C$_2$ and C$_3$ contributions to higher binding energy, while the enhancement around $E_3$ can be attributed to the carbonyl C$_1$ contribution of the molecule excited in the Rydberg state. 
The slight increase of signal around $E_2$ could be explained as well, since now two carbon 1s sites (C$_2$ and C$_3$) are expected to contribute at this energy for a photoemission taking place from the 3s Rydberg state, compared to only one contribution (C$_1$) from the ground state.

\subsection{\label{sec:sec:PECD} Time-resolved PECD}

\begin{figure*}[!ht]
	\centering
	\includegraphics[width = 1\textwidth]{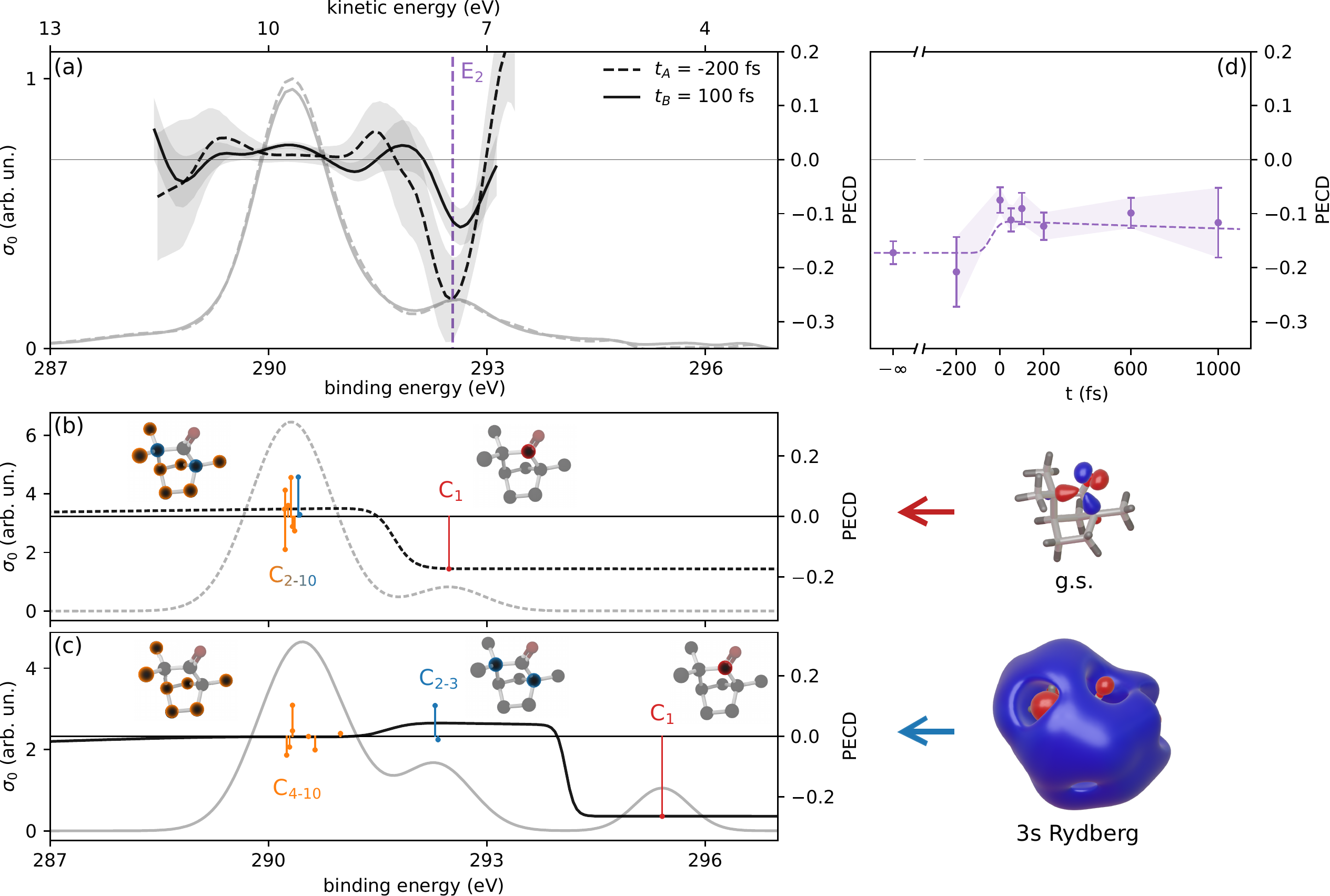}
	\caption{(a) Experimental PECD (black lines) and PES (gray lines) at $t_A = -200\,$fs (dashed lines), and $t_B=100\,$fs (solid lines), and corresponding $1\sigma$ PECD confidence intervals (shaded gray area), for S-(+)-Fenchone. Energy $E_2$ is indicated by a vertical violet dashed line. (b,c) Calculated PECD (S-(+)-Fenchone) for the electrons emitted from the Carbon-1s site from a  ground electronic state (b) and from excited 3s Rydberg state (c), indicated by red (C$_1$), blue (C$_{2}$, C$_3$) and orange (from C$_{4}$ to C$_{10}$) lollipops.  Total PECD in (b) and (c) shown with dashed and dotted black line, respectively. The theoretical PES is also drawn for reference (gray lines). (d) Measured TR-PECD at $E_2 =292.53\,$eV, shown with violet dots along with the 1$\sigma$ error-bars. The static value is reported at $t=-\infty$. The theoretical PECD from the model is shown as a dashed line overlaid to the experimental data. }
	\label{fig:3}
\end{figure*}

TR-PECD allows unveiling the chiral electronic structure and molecular relaxation dynamics after excitation by the visible laser. Figure \ref{fig:3}a compares the PECD measured at $t_A$ and $t_B$, represented with black dashed and solid lines, respectively. Since PECD is obtained as the dichroism (difference) signal normalised by the total signal, $\sigma_0$, the PECD becomes overwhelmed by noise as the PES intensity vanishes. Hence we only show PECD results for binding energy ranges where the PES intensity is sufficiently large to guarantee the reliability of the measurement. 
This, unfortunately, causes the exclusion of PECD measured in the E$_3$ energy range that spans the C$_1$ contribution in the Rydberg excited state.
 
A clear increase of the PECD around $E_2$ can be observed at $t_B$ compared to its value at $t_A$. This increase at positive pump-probe delay, corresponding to a decrease in absolute value of the PECD, can be interpreted as due to the different chiral asymmetry and binding energies of the 1s carbon states of the molecule in the excited 3s Rydberg state, that originates from the different local chiral environment of each carbon atom after excitation. For explaining this experimental observation, the theoretical PECDs of the photoelectrons coming from all the 1s carbon states of the molecule in the ground and excited 3s Rydberg state are shown with vertical lollipops in Figs. \ref{fig:3}(b) and \ref{fig:3}(c), respectively. The predicted total PECD, taking into account the resolution of the experiment, is overlaid in black.

The theoretical PECD at E$_1$ binding energy changes by very little upon the $n \rightarrow 3s$ excitation. Hence, we should not expect to observe a transient PECD response with high contrast, even if the ground state population were to be fully transferred to the Rydberg state.
In contrast, the experimentally observed PECD change at $E_2$ is clearly significant. Theoretically, the ground state PECD at $E_2$ is predicted to be negative, arising solely from the C$_1$ atom (see Sec. \ref{sec:staticPECD}). For the excited 3s Rydberg state, the C$_1$ atom no longer contributes at $E_2$ (having been up-shifted to $E_3$) and the net PECD at $E_2$ is predicted  to be positive, due to a positive C$_3$ contribution ($\sim\,0.1$ predicted) and an approximately zero C$_2$ contribution. In practice, assuming an initial molecular excitation of $\sim$12.5$\,\%$ in the experiment, the population in the excited Rydberg state at 100$\,$fs is predicted to be $\sim12\%$, and the reduced PECD magnitude observed at $E_2$ can thus be explained as the mean of $\sim12\%$ C$_2$ and $\sim$12$\%$ C$_3$ 3s Rydberg state contributions combined with the residual $\sim$88$\%$ ground state C$_1$ contribution.

The average PECD in correspondence of the peak at $E_2$ as a function of the pump-probe delay $t$, $\langle \mathrm{PECD} \rangle_2(t)$,  is shown in Fig. \ref{fig:3}(d) with violet dots.  For all positive delays the measured PECD is higher than the PECD at $t=-200\,$fs and the static PECD. 
We can compare the experimental PECD with the expected PECD at $E_2$ estimated by assuming a 12.5\% excitation probability and considering only variations caused by the different population of the ground and excited states as a function of the pump-probe delay. The computed PECD is overlaid to the experimental data and shown as a violet dashed line in Fig. \ref{fig:3}(d). The model predicts a step-like increase of the PECD, that slowly decreases towards the initial value with a time decay constant of 3.3$\,$ps. 
This predicted behavior is consistent to what is observed experimentally, and confirms the picture previously discussed for delay $t_B=100\,$fs. For all positive delays, the observed PECD can be then understood as the combination of the C$_2$ and C$_3$ 3s Rydberg state contributions with the residual ground state $C_1$ contribution. At $t_C = 50\,$fs, where the population in the excited Rydberg state reaches its maximum $\sim12.1\%$, the predicted PECD is $-11.47\%$, $5.81\%$ higher than the static value, and the measured PECD is $-11.2\pm2.2\%$. 
Compared with the simplified model proposed, the experiment may exhibit an additional dynamics on the ps time scale that could reflect the changes in the molecular geometry during the relaxation dynamics, as discussed by Beaulieu et al. \cite{beaulieu2016probing}, and/or the effects of alignment from the pump pulse that have here been neglected. Furthermore, the slight increase of the PECD at 0 fs and 100 fs compared to our model could be related to a fast vibrational dynamics that would require additional investigation.
Overall this TR-PECD measurement performed in the excited state allowed accessing the local chirality of the electronic environment around those sites that in the ground state are mixed with the contributions from all other sites. The combination of a site-selective probe and the chemical shift triggered by excitation with an ultrafast pump allowed us to measure the chiral structure of a molecule from partially-individual sites that, otherwise, would not be accessible. 

\section{\label{sec:conclusion} Conclusions and Perspectives}
In this work, we demonstrated core-level PECD at the carbon K-edge with FEL radiation, obtaining good agreement with theoretical CMS-X$\alpha$ calculations and previously reported synchrotron experiments. The core probe was exploited for studying the electronic scattering potential of photo-excited fenchone molecules. The photoelectron dichroism from the excited state was studied at different pump-probe delays and photoelectron kinetic energies, and the role of different sites of the ground state and excited molecule in the total measured PECD was unraveled. More specifically, the pump-excitation allowed control of the core-shell ionization site by isolating an additional subset of C-atoms (C$_2$ and C$_3$) whose PECD contribution could be unraveled, and whose dynamics could be rationalized with the help of high level PECD calculations which are highly predictive for core shells.

In the context of PECD, this work demonstrates explicitly its clear dependence to the initial orbital, in terms of localization within the molecule, and permits a comparison of the dichroism response at one of the asymmetrically substituted chiral centres, C$_3$, with that of adjacent C atom sites. The different chemical shifts of binding energy induced by the pump laser are particularly helpful for probing at different sites, opening the concept of advanced TR-chiral-XPS. Previous work on valence-shell PECD had shown a dependence with the nature and symmetry of the initial orbital but could not directly address the localization issue. In addition, we show the strong sensitivity of PECD to a change in the surrounding electronic density and not only to the molecular structures (geometries) as already reported \cite{nahon2015valence}.

The understanding of chiral dynamics that can be gained in this type of experiment, by following the complex ultrafast dynamics taking place in excited chiral molecule or during the dissociation of the molecule itself, will give new insight into fundamental questions in photochemistry and biochemistry related to the nature of chirality and to chiral recognition. Note that such a methodology may have a broad range of interests since it could be applied to any type of chirality, beyond the asymmetric-carbon central chirality case \cite{darquie2021valence}, as well as beyond the gas phase since PECD has been recently demonstrated in liquid fenchone \cite{pohl2022photoelectron}.

\begin{acknowledgements}
This project has received funding from the European Union’s Horizon
2020 research and innovation programme under Grant Agreement Nos. 674960 (ASPIRE), 860553 (SMART-X), 682978 (EXCITERS) and 654148 (LASERLAB-EUROPE), from the Italian Ministry of Research and Education with the projects ELI ESFRI Roadmap, from the Consiglio Nazionale delle Ricerche with the Joint Laboratory ATTOBIO. FC, MG and LC acknowledge support from the Deutsche Forschungsgemeinschaft (DFG, German Research Foundation) – SFB-925 – project 170620586 and the Cluster of Excellence Advanced Imaging of Matter (AIM) 
\end{acknowledgements}

\section*{Author contributions}
M.D., Y.M., M.N., and C.V. proposed the experiment. M.D., D.F., M.C., and S.T. performed preliminary experiment. 
C.C., M.D., G.DN, M.DF., D.F., G.A.G., O.P., and D.Y., executed the experiment, collected the experimental data and provided online analysis of the experimental data. 
S.B., V.B., D.C., E.C., A.G.C., L.C., M.C., G.C., M.G., L.N., M.N., I.P., P.P.G., K.C.P., A.P., S.S., S.T., K.U., C.V., and N.Z. participated in the experiment and discussed the results. M.D. and D.F. analyzed the experimental data. N.B. and I.P. performed the simulations. V.B., M.D., D.F., L.N., I.P. and C.V. interpreted the data and wrote the paper.  
All the authors discussed the results and contributed to the final manuscript.

{ 
\appendix
\section{\label{sec:methods:expset} Experimental Setup}
Experiments were performed at the seeded-FEL user facility FERMI, providing left and right circularly polarized FEL pulses with a high degree of polarization, typically greater than 95\% \cite{roussel2017polarization}, at 10 Hz repetition rate. A double-stage high-gain harmonic generation scheme allows to reach FEL wavelengths down to 4 nm, where the output of a shorter replica of a first undulator chain (FEL-1) is used to seed a second undulator chain (FEL-2). The seed laser is produced by an optical parametric amplifier pumped by a Ti:Sapphire laser. Its wavelength is set to $\lambda_{\mathrm{seed}} = 264.5\,$nm, and the 64$^{\mathrm{th}}$ harmonic is produced from the double cascade scheme, generating FEL probe pulses at 300$\,$eV photon energy. The FWHM pulse duration at this photon energy is estimated to be ~$15\,$ fs \cite{finetti2017pulse}.  The choice of 300$\,$eV is constrained by the asymmetry response of fenchone at the C 1s edge, the background contribution for the pump alone (especially the first ATI peak) and the available FEL photon energy in this range, which are given by the proper combination of harmonics in the two undulator chains.

The seeded scheme allows pump-probe experiments with a timing jitter below 10$\,$fs \cite{cinquegrana2014optical, lyamayev2013modular}. A portion of the infrared (IR) laser that drives the seed laser of the FEL is propagated to the experimental hall by a high-stability optical beam transport to seed also the UV pump laser. Second harmonic generation inside a BBO crystal generates 100 $\mu$J pump pulses with 55--75 fs pulse duration at 396$\,$nm. The intensity is then attenuated to 31 $\mu$J by using a half-waveplate and a polarizer to reduce photoionization by the pump alone and suppress sidebands in the photoelectron spectrum arising from the dressing of the continuum electrons by the UV laser photons. The pump and probe pulses are then recombined at the interaction region in a quasi-collinear geometry, at an angle of 0.6°, and the temporal and spatial overlap is finely optimized by means of well-tested spectroscopy methods in noble state gases \cite{finetti2017optical}. The probe is circularly polarized and propagates along the axis $z$, while the pump is linearly polarized along the axis $y$ perpendicular to $z$. Both axis $y$ and $z$ are parallel to the plane of the VMI detector placed on top.

The molecular sample of S-(+)-Fenchone (Aldrich, 98\%) was placed in an external reservoir and heated up to 85°C. The fenchone vapor was then mixed with 1 bar of helium that serves as a carrier gas. A pulsed Parker Valve with a nozzle diameter of 800$\mu m$ generated a supersonic molecular gas jet that was then collimated by a skimmer with a diameter of 1\,mm, placed at a distance of 40 cm from the interaction zone. No clusters of fenchone have been detected to be formed in such seeded conditions.

The VMI static electric fields were adjusted in order to collect photoelectrons up to $\sim$18 eV kinetic energy. The VMI is equipped with a pair of resistance matched MCPs (chevron double-stack MCP) coupled to a phosphor screen, with an active diameter of 77\,mm. The electron images are then recorded with a CCD camera system. 

\section{\label{sec:methods:data} Data acquisition and analysis}
\subsection{\label{sec:methods:data1} Static experiment}
The \emph{static} FEL-only PECD experiment was performed by alternatively recording VMI electron images from right and left circularly polarized FEL radiation, switching the helicity every 10$\times$10$^3$ single-shot acquisitions. A \emph{background image} was acquired every 3 single-shot acquisitions by properly delaying the pulsed valve in order to collect the electrons coming only from the residual gas. A total of 200$\times10^3$ single-shot images were acquired, out of which one third are background images and the other are \emph{signal images}, corresponding to a total of 6 hours of acquisition. Background images were not subtracted from the signal data, since we used a background retrieval algorithm to reconstruct the actual background present in the signal, as discussed later.

The FEL intensity fluctuations were monitored by measuring the photocurrent (PC) coming from one of the plane mirrors used to transport the FEL radiation to the experimental chamber. The variation of this signal during all the acquisitions was measured to be of the order of $\sim\,8\%$. The 200$\times10^3$ original dataset was initially filtered to remove all the shots where the PC was lower than $40\%$ of its mean value. This procedure resulted in approximately $M\approx60\times10^3$ images for each helicity. 
Each single-shot image (of size $980\times980$ pixels) was processed by a centroiding algorithm in order to retrieve the positions of all the spots in the image corresponding to single electrons impinging in the MCP plate that were subsequently amplified by electron cascade inside the double-stack MCP. From the coordinates of each spot, images were reconstructed by 2D binning the point-based dataset into a $512\times512$ square grid of axes $y$ and $z$ covering the original image extension. Each image was convoluted with a 15$\times$15 pixels 2D gaussian kernel having $2.6$ pixels standard deviation in both $y$ and $z$ directions. This convolution parameter was chosen as a compromise between the need to reduce the shot (Poisson) noise arising from the low number of events detected and at the same time maintain an energy resolution sufficient enough to distinguish the two main peaks in the ground state XPS.

The dataset of $M$ images obtained for each helicity was then divided into $N=30$ sub-sets of equal dimension ($\sim\,2000$ images \emph{per} subset and helicity), and the images from each subset were averaged to provide $N$ pairs of left-handed $I^{+1}(y,z)$ and right-handed $I^{-1}(y,z)$ circularly polarized measurements.

From the acquired images, the original 3D photoelectron distribution can be retrieved from standard inversion techniques \cite{garcia2004two}. In the \emph{static} experiment of Fig. \ref{fig:1}, the PAD exhibits cylindrical symmetry around the axis $z$, and can be defined as the following Legendre polynomial expansion:
\begin{equation}
\mathrm{PAD}^p(K,\theta) = \sum_{l=0}^2 B^p_l(K) P_l(\cos\theta),
\label{eq:1}
\end{equation}
where $P_l$ are the Legendre polynomials of order $l$, which depend on the polar angle $\theta$ with respect to the axis $z$, $B_l^p$ are the anisotropy parameters, expressed as a function of the kinetic energy $K$ of the outgoing electron, and $p$ is the helicity. Frequently it is convenient to use the normalized anisotropy parameters $b_l^p=B^p_l/B^p_0$. For even $l$, $b_l = b_l^{+1} = b_l^{-1}$, while for odd $l$, $b_l^{+1} = - b_l^{-1}$. The VMI records the projection of this 3D distribution on the 2D plane of cartesian axis $y$ and $z$. To reconstruct the $b_l$ parameters from this projection, the acquired VMI images were processed using the cpBasex code \cite{champenoiscpbasex}, which implements the pBasex inversion method \cite{garcia2004two} without polar rebinning.

Due to the cylindrical symmetry of the 3D distribution and the rules for the coefficients $b_l$, 2D images should obey the two symmetry properties: $I^{p}(y,z) = I^{-p}(y,-z)$ and $I^{p}(y,z) = I^{p}(-y,z)$, for $p=\pm1$. Before performing the inversion, we further aggregated our data taking into account these symmetries, in order to increase the signal-to-noise and at the same time compensate for the unavoidable non-uniformity of the imaging detector. A single dataset of $N$ symmetrized images $S$ was constructed by the symmetrization procedure:
$
S(y,z) = [{I^{+1}(y,z) + I^{+1}(-y,z) + I^{-1}(y,-z) + I^{-1}(-y,-z)}]/{4},
$
and Abel inverted in order to retrieve, for each sample, the three anisotropy parameters $B_{0}(K)$, $B_{1}^{+1}(K)$ and $B_{2}(K)$.

The retrieved coefficient $B_{0}(K)$ exhibited two clear peaks, along with a broad background signal that was stronger than the one measured every three shots of the laser with the pulsed valve delayed, but similar in shape.
This background likely comes from the photoionization from the valence states of the residual gas and the helium carrier by the XUV pulse. Due to their high kinetic energy (greater than 250 eV), these electrons were not collected by the VMI, but are expected to hit the electrodes and walls of the VMI generating secondary electrons in a random direction, part of which are then accelerated towards the detector. Because of the stochastic nature of this secondary emission and the achiral nature of helium, this background is not expected to change with the helicity of the laser. 
As such, it is expected to only add up to the reconstructed even anisotropy parameters of Eq. \ref{eq:1}.  By using a background retrieval algorithm described in the Supplemental Material \cite{supplemental}, $B_0(K)$ is expressed as the sum of a signal $B_0^{sig}(K)$ and a background $B_0^{bg}(K)$. See Figure \ref{fig:S1} in the Supplemental Material \cite{supplemental}. 
The procedure is repeated for each of the $N=30$ samples
and the final PES $\sigma_0(K)$ is computed from the mean of the $N$ retrieved $B_0^{sig}(K)$, while PECD$(K) = 2 b_1^{+1}(K)$ is computed from the mean of the $N$ normalized asymmetries $b^{+1}_1(K)={B_1^{+1}(K)}/{B_0^{sig}(K)}$. 
The 1$\sigma$ confidence intervals of $\sigma_0(K)$ and PECD$(K)$ are computed from the standard deviation of the mean estimated from the standard deviation of the $N$ samples. These results, that can be conveniently expressed both as a function of the kinetic energy $K$ and binding energy $E$, are shown in Figs. \ref{fig:1}(a,c). The PECD associated to a peak in the PES centered at the binding energy E$_{j}$ is estimated as the weighted average:
\begin{equation}
    \langle \mathrm{PECD} \rangle_j = \frac{\int_{E_j - \Delta E} ^{E_j + \Delta E} G_{\alpha_g}(E-E_j) 2 b_1^{+1}(E) \mathrm{d}E}{\int_{E_j - \Delta E} ^{E_j + \Delta E}G_{\alpha_g}(E-E_j)\mathrm{d}E} 
\end{equation}
where $\Delta E = 0.4\,$eV and $G_{\alpha_g}$ is the one-electron Gaussian response function derived from the fit described in Sec. \ref{sec:methods:fitting}. The estimate is computed for each of the $N$ normalized asymmetries $b^{+1}_1$, from which the mean and 1$\sigma$ confidence interval are extracted, shown in Figs. \ref{fig:1}(e,f) with squares and errorbars, respectively.

\subsection{\label{sec:methods:data2} Time-resolved measurement}

In the delayed pump-probe experiments the UV pump pulse, linearly polarized along the direction $y$, may induce an alignment of the molecule, breaking the cylindrical symmetry around the FEL propagation axis $z$. In principle, this invalidates use of the Abel inversion procedure, requiring such axial symmetry, for the direct extraction of the coefficients of Eq. \ref{eq:1}. Without assuming this symmetry the photoelectron angular distribution can be generally written as an expansion of spherical harmonics, $Y_l^m(\theta,\varphi)$:
\begin{equation}
I^p(K,\theta, \varphi) = \sum_{l=0}^{2n} A^p_{l,m}(K) Y_l^m(\theta,\varphi)
\end{equation}
with the summation running up to $l=2n$, where $n$ is the number of photons in the combined excitation/ionization process (here $n=3$). The breaking of the cylindrical symmetry can result in expansion coefficients $A_{l,m}\ne 0$ for $m\ne0$. As before it can be convenient to introduce the normalized angular coefficients $a_{lm}=A_{lm}/A_{00}$.

To further explore this issue we computed theoretical 3D PADs including the broken cylindrical symmetry caused by the crossed pump-probe polarizations. For the carbonyl C$_1\ 1s^{-1}$ photoelectron angular distribution from the two-photon excited fenchone 3s Rydberg at 300 eV photon, we obtained $a^{(+1)}_{10} = -0.074$, $a^{(\pm1)}_{22}=0.027$  but $|a_{l,m}|\ll 0.02$ for all the other harmonics (Table \ref{tab:3dpad} and Fig. \ref{fig:almbars}; further detail in Sec. \ref{sec:methods:theory} below). Neglecting these small harmonic terms, and noting the relation $Y_{l0}(\theta,\phi)=\sqrt{\frac{2l+1}{4\pi}}P_l(\cos \theta)$, the dominant chiral coefficient, $a^{(+1)}_{10}$, expressed for an alternative PAD expansion in Legendre polynomials becomes $b^{(+1)}_1=\sqrt{3}a^{(+1)}_{10}=-0.128$. This agrees closely with the value $b^{(+1)}_1=-0.129$ that is computed for an assumed random orientation (shown in Fig. \ref{fig:3}c). Examination of the $a^{(+1)}_{lm}$ coefficients for the C$_3$ 1s$^{-1}$ ionization (Table \ref{tab:3dpadC3}) leads to a similar conclusion. Hence, in these instances the presence of harmonics $l  > 2$ in the two-photon excited PAD and its broken axial symmetry will have little influence on our theoretically calculated PECD asymmetry.

\begin{table*}
	\begin{tabular}{c | c@{\hskip 0.58 cm} c@{\hskip 0.58 cm} c@{\hskip 0.58 cm} c@{\hskip 0.58 cm} c@{\hskip 0.58 cm} c@{\hskip 0.58 cm} c@{\hskip 0.58 cm} c@{\hskip 0.58 cm} c@{\hskip 0.58 cm} c@{\hskip 0.58 cm} c@{\hskip 0.58 cm} c@{\hskip 0.58 cm} c}
     $m=$& -6 & -5 & -4 & -3 & -2 & -1& 0& 1& 2& 3 & 4 & 5 & 6\\
\colrule
 $l=$0& &  &  &  &  &  & 1 &  &  &  &  &  &\\
 $l=$1&  &   &   &   &   & 0 & -0.0739 & 0 &   &   &   &   &\\
 $l=$2&  &   &   &   & 0.0181 & 0 & 0.0077 & 0 & 0.0268 &   &   &   &\\
 $l=$3&  &   &   & 0 & -0.0147 & 0 & 0.0057 & 0 & 0.0074 & 0 &   &   &\\
 $l=$4&  &   & -0.0134 & 0 & -0.0051 & 0 & -0.0010 & 0 & 0.0020 & 0 & 0.0168 &   &\\
 $l=$5&  & 0 & -0.0008 & 0 & -0.0003 & 0 & -0.0054 & 0 & -0.0074 & 0 & -0.0064 & 0 & \\
 $l=$6& 0 & 0 & 0 & 0 & 0 & 0 & 0.0034 & 0 & 0.0043 & 0 & 0.0030 & 0 & 0\\
 \colrule
	\end{tabular}
	\caption{\label{tab:3dpad}%
		Normalised coefficients, $a^{(+1)}_{lm}$, for the calculated S-(+)fenchone C$_1$ 1s$^{-1}$ PAD expanded in spherical harmonic functions, $Y_{lm}$. The calculation models the 3D photoelectron distribution, $I(\theta,\varphi)$, from the two-photon pump-aligned 3s Rydberg state that is ionized by a 300 $\,$eV, LCP probe pulse.}
\end{table*}

\begin{figure}
	\centering
    \includegraphics[width =0.5\textwidth]{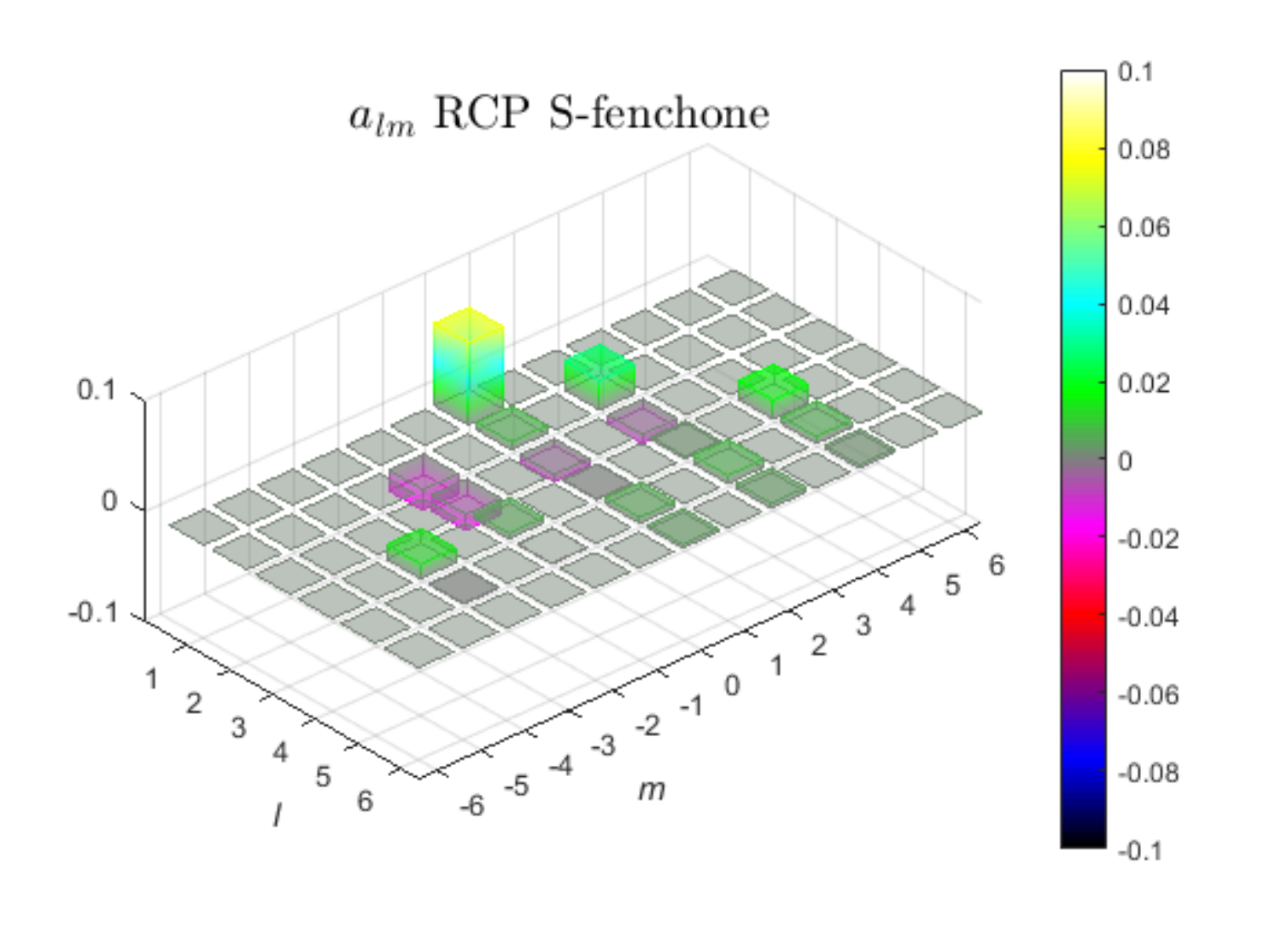}
	\caption{\label{fig:almbars} The $a_{lm}$ coefficients, $l=1$ to 6, for the calculated S-(+)fenchone C$_1$ 1s$^{-1}$ PAD. See Table \ref{tab:3dpad} and note that this figure is for RCP polarization ($p=-1$).}
\end{figure}

\begin{table*}
	\begin{tabular}{c | c@{\hskip 0.58 cm} c@{\hskip 0.58 cm} c@{\hskip 0.58 cm} c@{\hskip 0.58 cm} c@{\hskip 0.58 cm} c@{\hskip 0.58 cm} c@{\hskip 0.58 cm} c@{\hskip 0.58 cm} c@{\hskip 0.58 cm} c@{\hskip 0.58 cm} c@{\hskip 0.58 cm} c@{\hskip 0.58 cm} c}
     $m=$& -6 & -5 & -4 & -3 & -2 & -1& 0& 1& 2& 3 & 4 & 5 & 6\\
\colrule
 $l=$0& &  &  &  &  &  & 1 &  &  &  &  &  &\\
 $l=$1&  &   &   &   &   & 0 & 0.0263 & 0 &   &   &   &   &\\
 $l=$2&  &   &   &   & -0.0320 & 0 & -0.1071 & 0 & -0.0369 &   &   &   &\\
 $l=$3&  &   &   & 0 & 0.0033 & 0 & -0.0023 & 0 & -0.0030 & 0 &   &   &\\
 $l=$4&  &   & 0.0388 & 0 & 0.0147 & 0 & 0.0169 & 0 & 0.0241 & 0 & 0.0273 &   &\\
 $l=$5&  & 0 & -0.0019 & 0 & -0.0011 & 0 & 0.0056 & 0 & 0.0077 & 0 & 0.0067 & 0 & \\
 $l=$6& 0 & 0 & 0 & 0 & 0 & 0 & -0.0045 & 0 & -0.0058 & 0 & -0.0039 & 0 & 0\\
 \colrule
	\end{tabular}
	\caption{\label{tab:3dpadC3}%
		Normalised coefficients, $a^{(+1)}_{lm}$, for the calculated S-(+)fenchone C$_3$ 1s$^{-1}$ PAD. Other details as for Table \ref{tab:3dpad}.}
\end{table*}

The consequences of applying an Abel inversion algorithm to an image from a non-axially symmetric PAD has been discussed by Comby \textit{et al.} \cite{comby2016relaxation} who deduce that \textit{only} the even $b_l$ terms retrieved from symmetrized images would be affected. In particular, the apparent $B_0^{sig}$ retrieved in these circumstances will not be strictly equivalent to the actual cross section $\sigma_0$. However, there will nevertheless be a secondary affect for the odd (chiral) coefficients, $b^{\pm1}_{l=odd}$, since these are normalised to $B_0^{sig}$. 

These considerations together suggest that effects of the intermediate's alignment (including introduction of the $l=3,5$ harmonics) can reasonably be neglected in our modelling and data treatments, but with the caveat that precision of the experimentally derived chiral asymmetry will be somewhat reduced.

For the \emph{pump-probe} experiment, a background image was acquired once every 7 single-shot acquisitions, and a total of 460$\times10^3$ single-shot images were acquired, distributed unevenly among the pump-probe delays, and corresponding to an acquisition time of 13 hours for the entire pump-probe scan. After filtering the original dataset, taking into account only signal images and removing images acquired at low FEL intensity, the number of samples for each helicity is
${\sim}9\times10^3$ at $t =-200\,$fs and $t =1000\,$fs, ${\sim}34\times10^3$ at $t =0\,$fs,  $t = 50\,$fs and $t =600\,$fs, and ${\sim}38\times10^3$ at $t = 100\,$fs and $t =200\,$fs. The same analysis described previously was performed, and $N$ symmetrized $S(y,z)$ images were retrieved for each pump-probe delay, from which the $B_l$ parameters were extracted. 

The first coefficient $B_0(K)$ exhibits again two strong peaks coming mainly to the photo-emission from the carbon 1s states of the molecule still in the ground state and a broad background signal. Furthermore, as shown on Fig. \ref{fig:S1} in the Supplemental Material \cite{supplemental}, a strong peak at $K  {\sim}0.7\,$eV, and a weaker peak at $K  {\sim}3.8\,$eV are visible, corresponding to the photo-emission from the HOMO state (${\sim}-8.6\,$eV) following the absorption of three and four pump photons, respectively. This is the pump alone PES signal that is achiral since resulting from ionisation by the linearly polarized UV laser. Both the background and multi-photon signal are expected to only add up to the reconstructed even asymmetry parameters. The background $B_0^{bg}(K)$ originating from the ionizing laser was estimated from an independent FEL-only measurement, and extracted following the same procedure used for the static experiment. The multi-photon signal $B_0^{mp}(K)$ from the pump was measured directly from an independent measurement performed with the pump only. Each of the $N$ $B_0$ values retrieved from the Abel inversion for each pump-probe delay can be expressed as the sum $B_0(K) = B_0^{sig}(K) + c_{bg}B_0^{bg}(K) + c_{mp} B_0^{mp}(K)$, where the coefficients $c_{bg}$ and $c_{mp}$ are found by minimizing $B_0^{sig}(K)$ in the region $K < 3.7\,$eV, where no photoionization signal from the carbon 1s states is expected. Once the signal $B_0^{sig}(K)$ is retrieved from the background subtraction procedure, $\sigma_0(K)$, PECD$(K)$, $\langle\mathrm{PECD}\rangle_j$, and their associated 1$\sigma$ confidence intervals, are retrieved in the same way as in the static case, for each pump-probe delay.

\section{\label{sec:methods:theory} Theoretical Computations}
Core 1s electron binding energies of fenchone in the ground and excited 3s Rydberg states have been computed by a DFT (PBE0/6-311++G**) $\Delta$SCF approach with corrections for relativistic effects \cite{besley2009, besley2021}. For the excited state XPS, a valence $\beta$-spin electron is excited and then a  $\beta$-spin core electron ionized. 

Photoionization dynamics calculations have been performed using the CMS-X$\alpha$ method which obtains the bound and continuum electron eigenfunctions in a one-electron potential. The potential is constructed by partitioning the molecule into atom-centred overlapping spherical regions, using the Slater X$\alpha$ local density approximation for the exchange contribution. The use of this independent electron, frozen core method for PECD calculations has been fully described \cite{powis2000photoelectron} and a successful application to predict the C 1s PECD of ground state fenchone has been reported \cite{ulrich2008giant}. Initially we took the preferred parametrization identified in this earlier work, later adapting it with increased atomic sphere radii but retaining all else. For the 3s Rydberg state we adopted an approach previously used for calculations on camphor \cite{lehmann2013} and difluorobenzene isomers \cite{staniforth2013}. Further information on the generation of the ground and 3s Rydberg state potential can be found in Supplemental Material \cite{supplemental}.

 The majority of these CMS-X$\alpha$ PAD calculations were made for an assumed random orientation of fenchone molecules. To justify the neglect of alignment induced in the 3s Rydberg state excitation, a limited number of further calculations were made to predict the full 3D PAD (see above Sec. \ref{sec:methods:data2} 
 and Supplemental Material \cite{supplemental}). Adopting methods introduced to predict PADs obtained in the resonance enhanced multiphoton ionization of difluorobenzene isomers \cite{staniforth2013}, pyrazine \cite{suzuki2012}, and PECD of camphor \cite{lehmann2013},  molecule frame PADs were calculated for the single photon ionization of the 3s state using the CMS-X$\alpha$ model as above. These molecule frame PADs were then averaged over the Euler angles representing lab frame molecular axis orientation, weighted by the orientation dependent transition strengths for populating the excited state. In the present case the two-photon transition moment tensor for 3s excitation, calculated by a  CAM-B3LYP/dAug-cc-pVDZ TD-DFT calculation \cite{singh2020}, was used to estimate the intermediate state alignment. More details are provided in Supplemental Material \cite{supplemental}.

\section{\label{sec:methods:fitting} Fitting and Modeling}
The static experimental PES was fitted by $\mathrm{PES}_{fit}(E)$, a series of 10 Gaussians, one for each $i$-th carbon atom, centered at the theoretical ground state ($g.s.$) binding energy $E_i^{g.s.}$:
\begin{eqnarray}
\mathrm{PES}_{fit}^{g.s.}(E) & = &  \sum_{i=1}^{10}{k_{i} G_{\alpha_g}(E - E_i^{g.s.})}; \nonumber\\
G_{\alpha_g}(x) & = & \frac{1}{\sigma_g\sqrt{2\pi}} e^{-x^2 / (2\sigma_g^2)} 
\label{eq:meth:1}
\end{eqnarray} 
where $\sigma_g$ is constrained to follow the resolution scaling law
\begin{equation}
    \sigma_{g} = \alpha_{g}\sqrt{K} = \alpha_{g}\sqrt{E-300\,\mathrm{eV}}\ ,
\label{eq:vmires}
\end{equation}
with the scaling factor $\alpha_{g}$. The coefficients $k_{2-10}$ are constrained to be equal, since the overlap in binding energy of the $C_{2-10}$ contributions doesn't allow to extract their independent values from the experiment, while $k_1$ is extracted from the fit. 
The fit returns a sigma resolution $\sigma_g = 0.5\,$eV at $K = 8\,$eV ($\alpha_{g} = 0.177 \sqrt{\mathrm{eV}})$, corresponding to a FWHM of 1.18 eV at this energy, and $k_1\approx1.36 k_{2-10}$. This resolution, lower than the one achievable with the VMI spectrometer available, comes mostly from the post-processing convolution of the VMI images already described in Sec. \ref{sec:methods:data1} and required for the reducing the noise due to the low number of counts. The difference between $k_1$ and $k_{2-10}$ can be attributed to the different cross-sections of these contributions. 
As discussed later, reliable cross-sections can be computed only for some atomic sites. For this reason, the theoretical photoelectron spectra shown in Figs \ref{fig:2}b,c are derived by using the retrieved fitted response function $G_{\alpha_g}$ and by assuming that all the carbon atoms contribute with the same cross section.
The PECD for the $g.s.$ shown in Fig \ref{fig:1}d was derived by weighting the theoretical PECD of each contribution ($[2b_1^{+1}]_i^{g.s.}$) with the retrieved response function:
\begin{equation}
\mathrm{PECD}^{g.s.}_{fit}(E) = \frac{\sum_{i=1}^{10} k_{i} G_{\alpha_g}(E-E_{g.s.}^i) [2b_1^{+1}]^{g.s.}_i}{\mathrm{PES}^{g.s.}_{fit}(E)} 
\end{equation}

The three lineouts $\hat\sigma_{0}^{(j)}(t)$ at $E_j = E_1$, $E_2$, $E_3$ (Fig. \ref{fig:2}e) were fitted with $\hat\sigma_{fit}^{(j)}(t) = \hat\sigma_0^{(j)}(t_A) +  \Delta\hat\sigma_{0,fit}^{(j)}(t)$, where $ \hat\Delta\sigma_{0,fit}^{(j)}(t)$ is a single exponential convoluted with a Gaussian:
\begin{eqnarray}
 \Delta\hat\sigma_{0,fit}^{(j)}(t) &=& h_j f(t) ; \nonumber\\
 f(t) &=&  \left[\frac{1}{\sqrt{2\pi}\sigma_t} e^{\frac{-[t-t_0]^2}{2\sigma_t^2}}\right] * \left[ e^{-\frac{t}{\tau}} H(t)\right]
\end{eqnarray}
where $H(t)$ is the Heaviside function, $\sigma_t = 32.8$fs is the measured cross-correlation of the pump and probe pulses, $\tau = 3.3\,$ps is the measured lifetime of the Rydberg 3s state \cite{comby2016relaxation}, while $h_1$, $h_2$, $h_3$, and $t_0$ are parameters of the fit. The cross-correlation of the two pulses has been measured from the sidebands in the photoelectron spectra of Fenchone appearing at high pump laser intensities ($\sim$150$\mu$J).  The fit is performed globally on the three lineouts by minimizing the sum of squared errors, where each line-out has been previously normalized to its amplitude $max(|\Delta \hat \sigma_0^{(j)}|)$. 
Confidence intervals of the fitted lineouts are obtained by repeating the procedure for the $N$ constructed data samples. 

Estimates of the percentage of excited molecules $r_0$ can be obtained from the fitted parameters $h_1$, $h_2$, and $h_3$, under the assumption that all the carbon atoms contribute with the same cross-sections both before and after excitation. Before excitation, 9 carbon atoms contribute to the main C peak at binding energy 290.3 eV while after excitation 7 carbon atoms contribute to this peak. Since all the lineouts have been normalized to the value of $\hat\sigma_0$ at $E_1$ and $t_A = -200\,$fs, we derive $h_1\approx -\frac{2}{9}r_{0}$, from which a first estimate of the excitation probability $r_{0}\approx16.7\%$ is obtained. Analogously, the positive increase of signal $h_3\approx\frac{1}{9}r_{0}$ around energy $E_3$ gives an estimated percentage of excited molecules $r_{0}\approx15.3\%$. Under the same approximations, and taking into account that in correspondence of the energy $E_2$ we have 2 carbon atoms contributing in the excited state and 1 carbon atom contributing in the ground state, we derive $h_2\approx\frac{1}{9}r_{0}$, that however gives a lower number of excited molecules $r_{0}\approx5.6\%$. Taking into account these considerations, we estimated a percentage of excitation $r_0\approx12.5\%$, obtained from the mean of the previous three estimates, associated to a standard error of the mean of $\sim3.5\%$. 
From these considerations, we assumed that after excitation $r_0= 12.5$\% of the sample is found in the excited 3s Rydberg state, while the rest $(1-r_0)$ is in the ground state. 

Since, after excitation, the excited Rydberg state relaxes to the ground state with a decay time $\tau = 3.3\,$ps, the percentage $r$ of the sample in the excited 3s Rydberg state changes with pump-probe delay $t$ as $
r(t) = r_0 f(t)$. 
The time dependent PES, which depends on both $t$ and $E$, formally originates from the overlap between these two contributions:
\begin{eqnarray}
\mathrm{PES}(E,t) = [1-r(t)] \mathrm{PES}^{g.s.}_{1-10}(E) + r(t) \mathrm{PES}^{3s}_{1-10}(E);\nonumber\\
\mathrm{PES}^{g.s., 3s}_{1-10}(E) = \sum_{i=1}^{10}{c_i^{g.s., 3s} G_{\alpha_g}(E - E_i^{g.s., 3s})};\quad\quad\quad
\end{eqnarray}
where $c_i^{g.s.}$ and $c_i^{3s}$ are the cross sections of the $i$-th Carbon state in the ground state and 3s Rydberg state, respectively, $E_i^{g.s.}$ and $E_i^{3s}$ are the binding energies of the $i$-th carbon state in the ground state and 3s Rydberg state, respectively, and the subscript $1-10$ indicates that all the atomic contributions from C$_1$ to C$_{10}$ are considered.
The time dependent PECD can also be formally written as the weighed average:
\begin{equation}
\mathrm{PECD}(E, t) = \frac{[1-r(t)] \mathrm{PECD}^{g.s.}_{1-10}(E)\mathrm{PES}^{g.s.}_{1-10}(E) + r(t) \mathrm{PECD}^{3s}_{1-10}(E) \mathrm{PES}^{3s}_{1-10}(E)}{[1-r(t)] \mathrm{PES}^{g.s.}_{1-10}(E) + r(t) \mathrm{PES}^{3s}_{1-10}(E)},
\end{equation}
where
\begin{equation}
\mathrm{PECD}^{g.s., 3s}_{1-10}(E) = \frac{\sum_{i=1}^{10} c_{i}^{g.s.,3s} G_{\alpha_g}(E-E_i^{g.s., 3s}) [2b_1^{+1}]^{g.s., 3s}_i}{\mathrm{PES}^{g.s., 3s}_{1-10}(E)}.
\end{equation}
$\mathrm{PES}(E,t)$ and $\mathrm{PECD}(E, t)$ could be in principle estimated by theoretically computing the binding energies, the cross-sections and asymmetry parameters for all the carbon contributions in the $g.s.$ and $3s$ Rydberg state. We computed the binding energies in the $g.s.$ and $3s$ Rydberg with Time-Dependent Density Functional Theory (TDDFT) calculations (see Table \ref{tab:table1} of the Supplemental Material \cite{supplemental}), and we computed the cross-sections and the asymmetry parameters with the CMS-X$\alpha$ method. However, the near-degeneracy in binding energy of several carbon atoms prevents to compute reliable theoretical cross-sections $c_i^{g.s., 3s}$ for atoms C$_2$-C$_{10}$ in the ground state and atoms C$_4$-C$_{10}$ in the excited 3s Rydberg state. 

For the Carbon 1s orbitals with better separated binding energies (like C$_1$ in the $g.s.$ and C$_1$, C$_2$, and C$_3$ in the 3s Rydberg state), a reliable approach that uses relaxed localization constraints has been implemented \cite{ulrich2008giant}. We used this approach for computing the cross sections $c_1^{g.s.}$ and $c_{1-3}^{3s}$ (listed in Table \ref{tab:table2} of the Supplemental Material \cite{supplemental}). The asymmetry parameters $[2b_1^{+1}]^{g.s., 3s}_i$ are not affected by these basis set constraints.

At the energy $E_2$, the PECD depends only on the cross sections and asymmetry parameters of the atoms C$_1$, C$_2$ and C$_3$. In this case, we can reliably compute the expected time-dependent PECD and compare it with the experimental one. The PECD at $E_2$ reads:
\begin{equation}
\mathrm{PECD}(E_2, t) \approx \frac{[1-r(t)] \mathrm{PECD}^{g.s.}_{1}(E_2)\mathrm{PES}^{g.s.}_{1}(E_2) + r(t) \mathrm{PECD}^{3s}_{2-3}(E_2) \mathrm{PES}^{3s}_{2-3}(E_2)}{[1-r(t)] \mathrm{PES}^{g.s.}_{1}(E_2) + r(t) \mathrm{PES}^{3s}_{2-3}(E_2)},
\end{equation}
and its value as a function of $t$ is shown with a dashed line in Fig. \ref{fig:3}(d).
The description can be further simplified with the approximation $E_{1}^{g.s.}\approx E_{2}^{3s} \approx E_{3}^{3s} \approx  E_2$, which gives:
\begin{equation}
\mathrm{PECD}(E_2, t) \approx \frac{[1-r(t)] c_1^{g.s.} [2b_1^{+1}]_1^{g.s.} + r(t)  \left\{c_2^{3s} [2b_1^{+1}]_2^{3s} + c_3^{3s} [2b_1^{+1}]_3^{3s}\right\}}{[1-r(t)]c_1^{g.s.} + r(t)[c_2^{3s} + c_3^{3s}]}
\end{equation}
CMS-X$\alpha$ calculations with relaxed localization constraints for C$_{1-3}$ predict c$_3^{3s}$/c$_1^{g.s.} \approx 1.67$, c$_2^{3s}$/c$_1^{g.s.} \approx 1.08$, $[2b_1^{+1}]_1^{g.s.}\approx-17.3\%$, $[2b_1^{+1}]_2^{3s}\approx-1.1\%$ and
$[2b_1^{+1}]_3^{3s}\approx10.2\%$. At the maximum of the excitation probability (around $t_C =50$ fs, where $r(t_C) \approx 12.1\%$), we calculate then:
\begin{equation}
\mathrm{PECD}(E_2, t_C) \approx -11.47\%
\end{equation}
with an increase in the PECD with respect to the ground state value of $5.81\%$, compatible to what has been experimentally observed on Fig. \ref{fig:3}d.
} 
\bibliography{bibliography}

\onecolumngrid
\pagebreak
\widetext
\begin{center}
\textbf{\large Supplemental Material \\ Time-resolved chiral X-Ray photoelectron spectroscopy with transiently enhanced atomic site-selectivity: a Free Electron Laser investigation of electronically excited fenchone enantiomers}
\end{center}
\setcounter{equation}{0}
\setcounter{figure}{0}
\setcounter{table}{0}
\setcounter{page}{1}
\setcounter{section}{0}
\renewcommand*{\thesection}{S.\the\value{section}}
\makeatletter
\renewcommand{\theequation}{S\arabic{equation}}
\renewcommand{\thefigure}{S\arabic{figure}}
\renewcommand{\thetable}{S\arabic{table}}

\section{Background subtraction}
\begin{figure}[!ht]
	\centering
	\includegraphics[width = 1\textwidth]{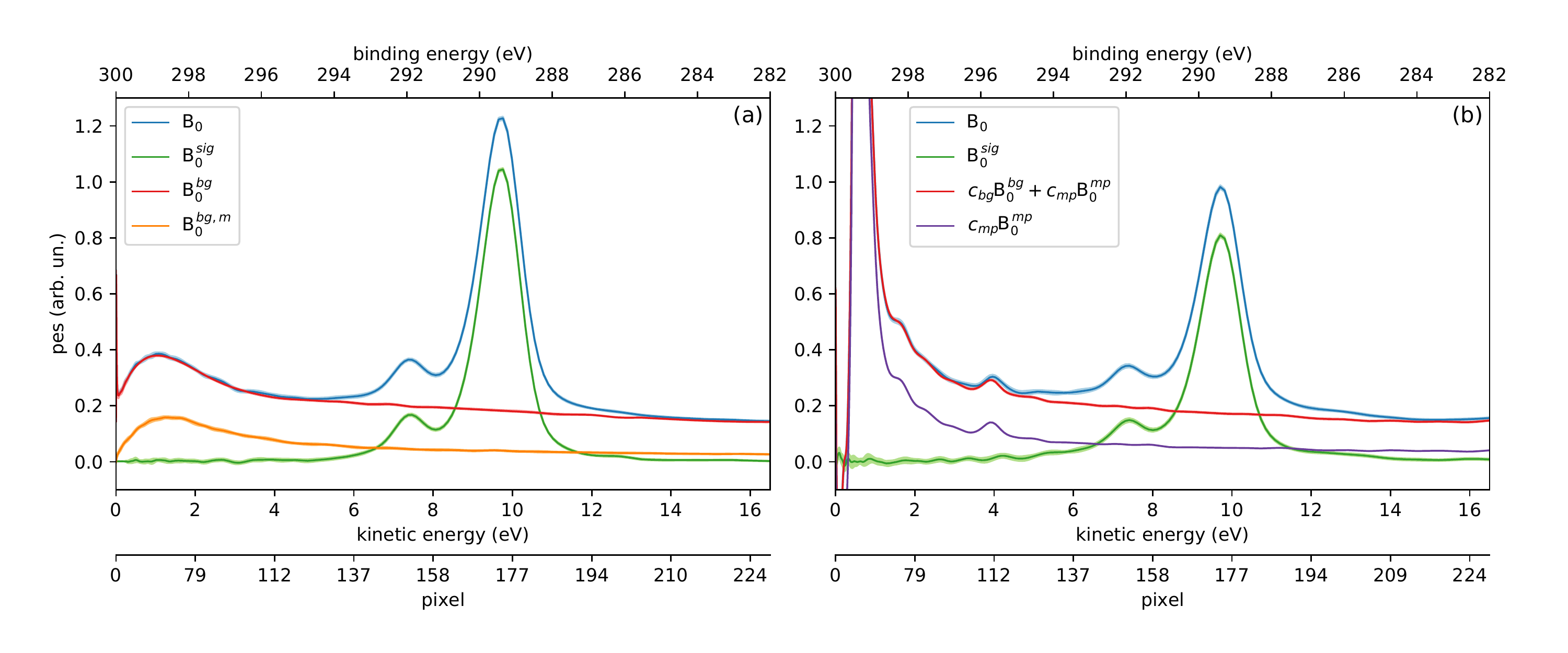}
	\caption{(a) Results of the background subtraction for the static experiment: measured $B_0$ (blue), measured background $B_0^{bg,m}$ (orange), retrieved background $B_0^{bg}$ (red), retrieved signal $B_0^{sig}$ (green). (b) Results of the background subtraction for the pump-probe experiment at $t = 0\,$fs: measured $B_0$ (blue), multi-photon ionization signal from the UV laser
	$c_{mp}B_0^{mp}$ (purple), retrieved total background signal $c_{bg}B_0^{bg} + c_{mp}B_0^{mp}$ (red), retrieved signal $B_0^{sig}$ (green). All the lines shown in (a) and (b) represent the average obtained from the $N$ samples acquired, and the shaded areas represent the associated 3$\sigma$ error.}
	\label{fig:S1}
\end{figure}

The first coefficient $B_0$ for the \emph{static} experiment is shown in blue in Fig. \ref{fig:S1}a. The background $B_0^{bg,m}$, measured by delaying the pulsed valve once every three shots of the FEL laser, is shown in orange in Fig. \ref{fig:S1}a. $B_0$ clearly exhibits a higher background, compared to the measured one. We can represent $B_0(r)$, function of the radial coordinate $r$, as the sum $B_0(r) = B_0^{sig}(r) +  B_0^{bg}(r)$, where $B_0^{sig}(r)$ is the signal and $B_0^{bg}(r)$ is the background. $B_0^{sig}(r)$ and $B_0^{bg}(r)$ are obtained from a background retrieval algorithm that is based on the main assumption that $B_0^{bg}(r)$ slowly varies with $r$. An initial estimate of the background  $B_0^{bg'}(r; A, C, \alpha_g)$, depending on three parameters  $A$, $C$, $\alpha_g$, is given by the following expression:
\begin{equation}
B_0^{bg'}(r; A, C, \alpha_g) =B_0(r) - \left[ A G_{\alpha_g}(K(r)-K_i) + C\sum_{i=2}^9 G_{\alpha_g}(K(r) - K_i)\right]
\end{equation}
where $K(r)$ is the calibration function that relates the kinetic energy $K$ to $r$ , $K_i$ is the theoretical kinetic energy of the emitted photoelectron of carbon C$_i$, and $G_{\alpha_g}$ is the Gaussian function defined in Eq. \ref{eq:meth:1} of the main text. The background $B_0^{bg}(r; A, C, \alpha_g)$ is obtained from $B_0^{bg'}(r; A, C, \alpha_g)$ by smoothing with a Savitzky-Golay filter (order = 3, window = 51 pixels). The optimum coefficients $A$, $C$, and $\alpha_g$ are then retrieved by minimizing, with the Nelder-Mead algorithm, the scalar function:
\begin{equation}
\epsilon(A, C, \alpha_g)= \sum_j |B_0^{bg}(r_{j+1};A, C, \alpha_g) - B_0^{bg}(r_{j};A, C, \alpha_g)|^2,
\end{equation}
where $r_{j}$ is the radius of the $j$-th pixel. Green and red lines in Fig. \ref{fig:S1}a represent the retrieved $B_0^{sig}$ and $B_0^{bg}$, respectively. The background retrieval procedure is applied to each of the $N$ samples. All the lines in Fig. \ref{fig:S1}a represent the average obtained from these $N$ samples, and the shaded areas represent the associated 3$\sigma$ error. 

For the \emph{pump-probe} experiment, the acquired $B_0$ for $t=0\,$fs is shown in blue in Fig.\ref{fig:S1}b, and the measured multi-photon ionization signal $B_0^{mp}$ coming only from the pump-laser is shown in purple. Three-photon and four-photon absorption of the UV light from the valence state of fenchone produce two additional peaks clearly visible in the low energy part of the spectrum and an additional background. 

The photoelectron spectrum associated to the photoionization from the carbon states by the FEL can be expressed as $B_0^{sig}(r) = B_0(r) - c_{bg}B_0^{bg}(r) - c_{mp} B_0^{mp}(r)$, where $B_0(r)$ is the measured signal, $B_0^{bg}(r)$ is the background retrieved from the static measurement and $B_0^{mp}(r)$ is the multi-photon ionization signal measured with the UV laser only.
The coefficients $c_{bg}$ and $c_{mp}$ are found by minimizing $\sum_j|B_0^{sig}(r_j)|^2$ in the region where no photoionization signal from the carbon 1s states is expected ($K<3.7\,$eV), using the Nelder-Mead minimization algorithm. Deviations of $c_{bg}$ and $c_{mp}$ from unity account for changes of the intensity of both signals due to laser intensity, gas density, and pressure fluctuations and drifts during the acquisition, as well as differences of the intensity of the two processes in the pump-probe experiment compared to the case where only the UV laser or the FEL laser are present. The retrieved $B_0^{sig}$ and $c_{bg}B_0^{bg}+c_{mp}B_0^{mp}$ for $t=0\,$fs are shown in Fig.\ref{fig:S1}b in green and red, respectively. The procedure is applied to each of the $N$ samples obtained for each pump-probe delay. All the lines in Fig.\ref{fig:S1}b represent the average obtained from the $N$ samples acquired at $t=0\,$fs, and the shaded areas represent the associated 3$\sigma$ error. 

\section{Experimental results}
\label{Experimental_results}
We summarize the experimental $\langle \mathrm{PECD}\rangle_2$ values obtained for the peak at energy $E_2 = 292.53\,$eV in table \ref{tab:exp}. 

\begin{table}[!ht]
	\begin{tabular}{c@{\hskip 0.07in}c@{\hskip 0.07in}c@{\hskip 0.07in}c@{\hskip 0.07in}c@{\hskip 0.07in}c@{\hskip 0.07in}c@{\hskip 0.07in}c@{\hskip 0.07in}c@{\hskip 0.07in}c}
	    &
		\textrm{Static}&
		\textrm{$t=-200\,$fs}&
		\textrm{$t=0\,$fs}&
		\textrm{$t=50\,$fs}&
		\textrm{$t=100\,$fs}&
		\textrm{$t=200\,$fs}&
		\textrm{$t=600\,$fs}&
		\textrm{$t=1000\,$fs}\\
		\colrule
		PECD ($\%$) &
		-17.2$\pm$2.1 & -20.8$\pm$6.5 & -7.5$\pm$2.4 & -11.2$\pm$2.2 & -9.1$\pm$2.9 & -12.3$\pm$2.5 & -9.9$\pm$2.8 & -11.7$\pm$6.4 \\
	\end{tabular}
	\caption{\label{tab:exp}%
		Measured PECD (in $\%$) at peak $E_2$, expressed as mean value $\pm$ 1$\sigma$ confidence interval, for the static case and time-resolved experiment.}
\end{table}

\section{Theoretical results}
\label{Theoretical_results}
TDDFT calculations (PBE0/6-311++G**) give two excited states, a LUMO state at $E = 4.29\,$eV corresponding to a n$\rightarrow\pi^*$ transition and a LUMO+1 state at $E = 5.81\,$eV corresponding to a n$\rightarrow$3s transition. With the CAM-B3LYP functional the transition energies are 4.35$\,$eV and 6.36$\,$eV, respectively. The binding energies computed using DFT-based $\Delta$SCF approach (PBE0/6-311++G**) for both the ground and excited states are reported in Table \ref{tab:table1}. 

\begin{table}
		\begin{tabular}{c@{\hskip 0.8in}c@{\hskip 0.8in}c@{\hskip 0.8in}c}
			\textrm{Carbon}&
			\textrm{$E^{g.s.}$, $g.s.$ (eV)}&
			\textrm{$E^{\pi^*}$, n$\rightarrow\pi^*$ (eV)} &
			\textrm{$E^{3s}$, n$\rightarrow3s$ (eV)}\\
			\colrule
			C$_1$ & 292.48 & 292.49 & 295.41\\
			C$_2$ & 290.43 & 290.72 & 292.33\\
			C$_3$ & 290.41 & 290.70 & 292.29\\
			C$_4$ & 290.31 & 290.30 & 290.99\\
			C$_5$ & 290.33 & 290.04 & 290.55\\
			C$_6$ & 290.22 & 290.11 & 290.64\\
			C$_7$ & 290.27 & 290.17 & 290.33\\
			C$_8$ & 290.26 & 290.32 & 290.29\\
			C$_9$ & 290.23 & 290.17 & 290.33\\
			C$_{10}$ & 290.23 & 290.06 & 290.25\\
		\end{tabular}
        \caption{\label{tab:table1}%
		Binding energies of carbon atom C 1s states for the ground state, and the excited n$\rightarrow\pi^*$ and n$\rightarrow3s$ states, computed using a DFT-based SCF approach (PBE0/6-311++G**) with a correction for relativistic effects.}
\end{table}
The CMS-X$\alpha$ photoionization dynamics calculation requires of a number of initial parameters to create the model potential. Here we take the preferred ground state potential parameters identified in earlier study of core ionization in fenchone \cite{ulrich2008giant}. This includes the fixed sphere radii, and a spherical harmonic basis of $l_{max}$ = 6/2/2 for, respectively, the asymptotic region of space, the first row atoms, and the hydrogen atoms. For continuum calculations, $L_{max}$\ limits are increased to 18/10/5, ensuring good convergence. To generate a 3s Rydberg state potential we adopted an approach previously used for calculations on camphor \cite{lehmann2013} and difluorobenzene isomers \cite{staniforth2013}. Starting from the converged ground state molecular potential a trial 3s Rydberg state potential was generated by promoting an electron from the HOMO into the 2$^{nd}$ virtual orbital, here of 3s character. These occupation numbers were frozen while the trial 3s state potential was re-iterated to achieve self-consistency.
 
 The preceding potential calculations were all repeated but using alternative atomic sphere radii generated by the Norman procedure \cite{Norman} scaled by an empirical 0.84 reduction factor, this having been generally been preferred in later PECD calculations. The Rydberg character of these alternative models is assessed by consideration of vertical excitation energies, changes in spatial extent $\Delta\left< r^2 \right>$ relative to the ground state and percentage asymptotic 's' character in Table \ref{tab:ip1}. Comparison with other \textit{ab initio} results \cite{singh2020} is also given. The Norman potential has slightly increased sphere radii, and from Table \ref{tab:ip1} was marginally preferred for the results presented here. In practice, however, little significant difference was found in the predicted PADs and cross-sections obtained from either choice of sphere radii.
 
 \begin{table}
	\begin{tabular}{c c c c c c}
		\textrm{}&
		\textrm{MS-X$\alpha$}\footnote{Using fixed atomic radii}&
		\textrm{MS-X$\alpha$}\footnote{Using Norman atomic radii$\times 0.84$ }&
		\textrm{TD-DFT}\footnote{Vertical excitation from TD-DFT CAM-B3LYP/dAug-cc-pVDZ calculation Ref. \cite{singh2020}} &
		\textrm{EOM-CCSD}\footnote{Vertical excitation from EOM-CCSD/cc-pVDZ+R calculations Ref. \cite{singh2020}} &
		\textrm{Expt.}\footnote{Spectroscopic origin Ref. \cite{baer1988}}\\
		\colrule
		$\Delta E\ (n \rightarrow 3s)$ eV & 5.1 & 6.0 & 6.26 & 6.21 & 5.952\\
		$\Delta\left<r^2\right>$ au & 23.1 & 20 & 51.7 & 64.5 &\\
		Rydberg orbital\footnote{Asymptotic composition} & 76\% s & 78.9\% s & &\\
	\end{tabular}
    \caption{\label{tab:ip1}%
		Computed properties of the 3s Rydberg state obtained by MS-X$\alpha$ calculation and comparison with TD-FT and EOM-CCSD results.}
\end{table}

\begin{figure}[!ht]
	\centering
	\includegraphics[width =0.35\textwidth]{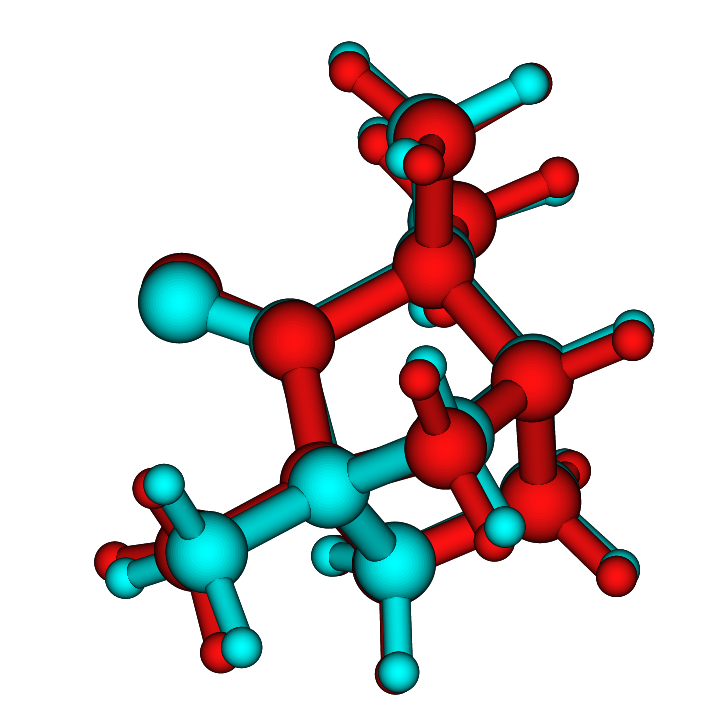}
	\caption{Comparison by a superposition of the ground state and 3s Rydberg state optimised equilibrium geometries (obtained from MP2/6-31G** and CAM-B3LYP/dAug-cc-pVDZ TD-DFT calculations).}
	\label{fig:ip1}
\end{figure}

All CMS-X$\alpha$ calculations, including those for the Rydberg state, were made with the original choice \cite{ulrich2008giant} of MP2/6-31G** optimized ground state molecular geometry. Little change to the equilibrium geometry is anticipated to result from the promotion of a nominally non-bonding (O lone pair) HOMO electron to a diffuse, non-bonding 3s Rydberg orbital. This was confirmed by separate TD-DFT calculations for the excited state equilibrium geometry, with a visual comparison given in Figure \ref{fig:ip1}.

 For both ground and excited 3s state potential calculations the angular basis of spherical harmonics used for computing the C 1s levels was restricted just to functions on the relevant atomic site, preventing any mixing of the 1s orbitals. This has the advantage of maintaining a clear localization of the 1s orbitals when nearly degenerate levels are considered (e.g.\ C$_4$ to C$_{10}$). But as previously described \cite{ulrich2008giant}, an alternative approach could be adopted when calculating the dipole matrix elements for those 1s orbitals with better separated Core Binding Energies (CBEs). In these cases the full molecular angular basis was used without imposing any restrictions, permitting possible mixing and polarization of the nominally spherical 1s electron density, while maintaining a clear atomic site localization. The calculations made with a restricted basis set to enforce strong localization, produce reduced cross-sections, presumably because overlap with the continuum function is more limited. However, PADs calculated with and without relaxing these basis set constraints vary imperceptibly. This can be understood as, due to being normalized by the total calculated cross-section ($b_j=B_j/B_0$), the angular distribution parameters depend only on the relative magnitudes of the matrix elements.

\begin{figure}[!ht]
	\centering
	\includegraphics[width =0.7\textwidth]{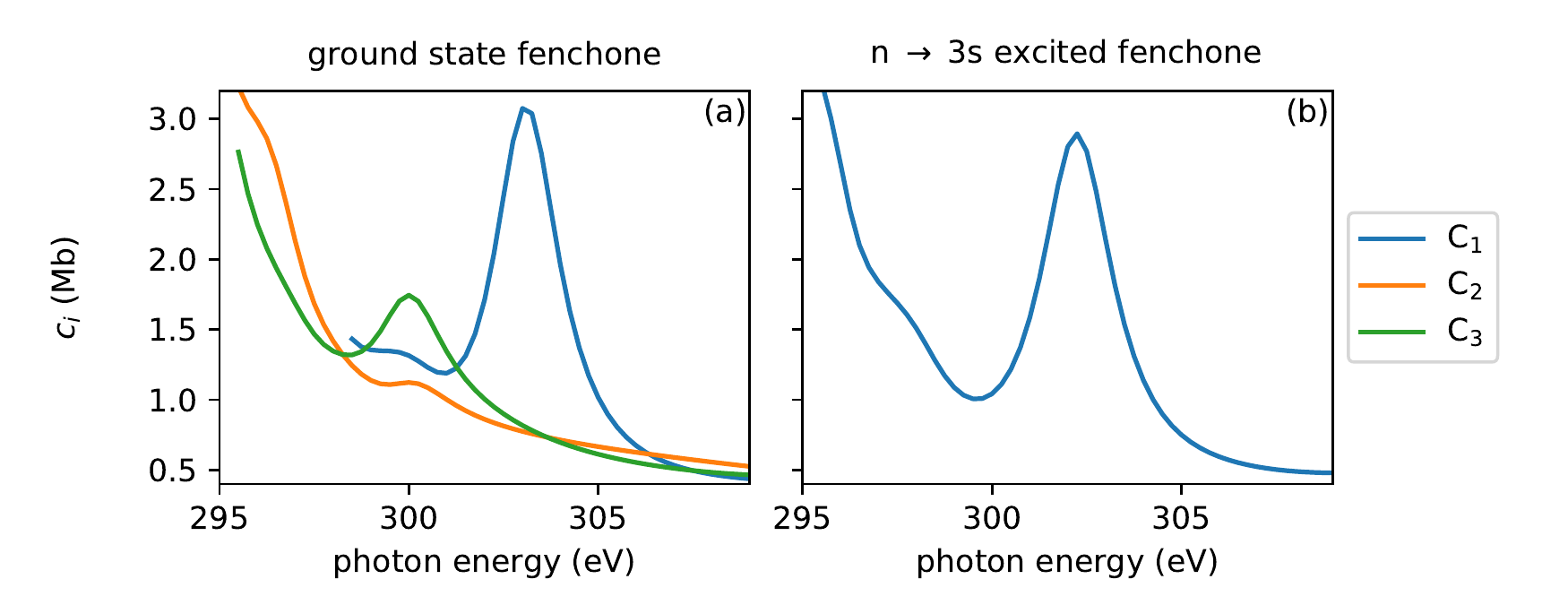}
	\caption{Cross sections $c_i$ for atoms C$_1$, C$_2$ and C$_3$ obtained with CMS-X$\alpha$ with relaxed constraints. The cross-sections are shown for ground state fenchone (a) and $n\rightarrow3s$  excited fenchone (b).}
	\label{fig:S2}
\end{figure}

\begin{figure}
	\centering
	\includegraphics[width =0.7\textwidth]{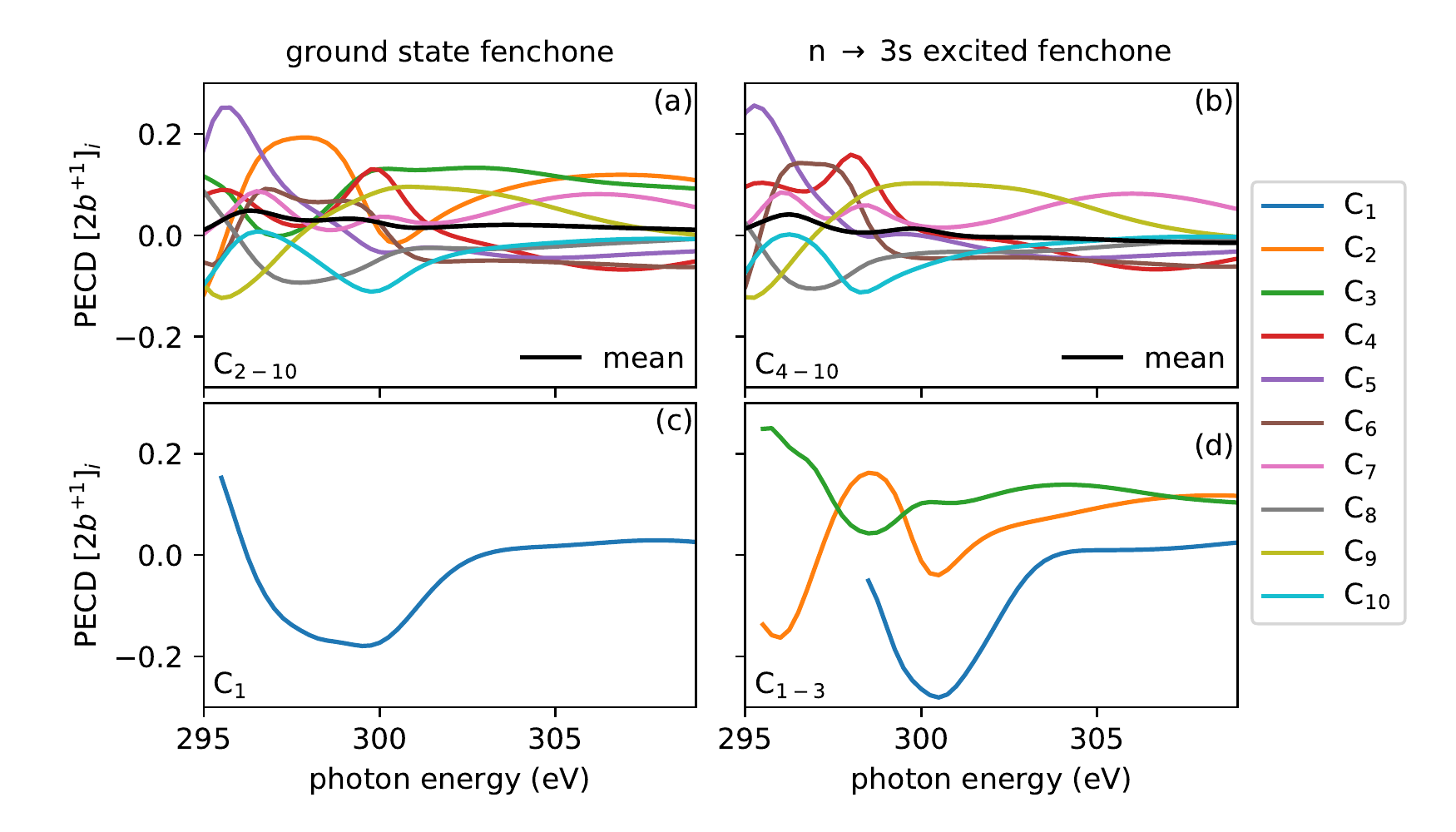}
	\caption{Core-level PECD $[2b_1^{+1}]_i$ of each atomic site  for each atom C$_i$ obtained with CMS-X$\alpha$ for Fenchone in the ground state (a,c) and $n\rightarrow3s$ excited Fenchone (b,d). }
	\label{fig:S3}
\end{figure}

Chiral asymmetry parameters for all the atoms have been computed by CMS-X$\alpha$ calculations using the Norman sphere radii potentials and maintaining strong atomic site localization of the core levels. For atoms C$_1$, C$_2$ and C$_3$ additional calculations were performed without imposing localization constraints. Figure \ref{fig:S2} shows, for different photon energies, the cross-sections obtained with this second approach for the molecule in the ground state and in the excited 3s Rydberg state. The cross-sections obtained at 300 eV are summarized in Table \ref{tab:table2}. We used these results to predict the relative weight of the ground and excited state to the observed PECD. Individual PECD values for each carbon atom are reported in Fig. \ref{fig:S3}. Table \ref{tab:table2} reports the $[2b_1^{+1}]_i$ values at 300$\,$ eV.

\begin{table}
	\begin{tabular}{c@{\hskip 0.3in}c@{\hskip 0.3in}c}
		\textrm{Carbon}&
		\textrm{$c^{g.s.}$, relax. (Mb)}&
		\textrm{$c^{3s}$,  relax. (Mb)}\\
		\colrule
		C$_1$ & 1.042 & 1.325 \\
		C$_2$ & - & 1.123  \\
		C$_3$ & - & 1.745  \\
	\end{tabular}
	\caption{\label{tab:table2}%
		Cross sections for the 300$\,$ eV photon energy ionization of ground state ($c^{g.s.}$) and excited n$\rightarrow3s$ state ($c^{3s}$) fenchone, computed using CMS-X$\alpha$, with relaxed constraints, permitting mixing (relax.). }
\end{table}

\begin{table}
	\begin{tabular}{c@{\hskip 0.3in}c@{\hskip 0.3in}c}
		\textrm{Carbon}&
		\textrm{$[2b^{+1}_1]^{g.s.}$}&
		\textrm{$[2b^{+1}_1]^{3s}$}\\
		\colrule
		C$_1$ & -0.173 & -0.264  \\
		C$_2$ & 0.007 & -0.011  \\
		C$_3$ & 0.131 & 0.102  \\
		C$_4$ & 0.129 & 0.009 \\
		C$_5$ & -0.033 & -0.0004  \\
		C$_6$ & 0.024&  -0.045    \\
		C$_7$ & 0.037 & 0.018   \\
		C$_8$ & -0.047 & -0.035   \\
		C$_9$ & 0.087 & 0.103  \\
		C$_{10}$ & -0.109 &-0.062   \\
	\end{tabular}
	\caption{\label{tab:3}%
		Core-level S-(+)-fenchone PECD $[2b^{+1}]_i$ for each atomic site C$_i$ obtained with CMS-X$\alpha$ at 300$\,$eV photon energy.}
\end{table}

\begin{figure}
	\centering
	\includegraphics[width =0.7\textwidth]{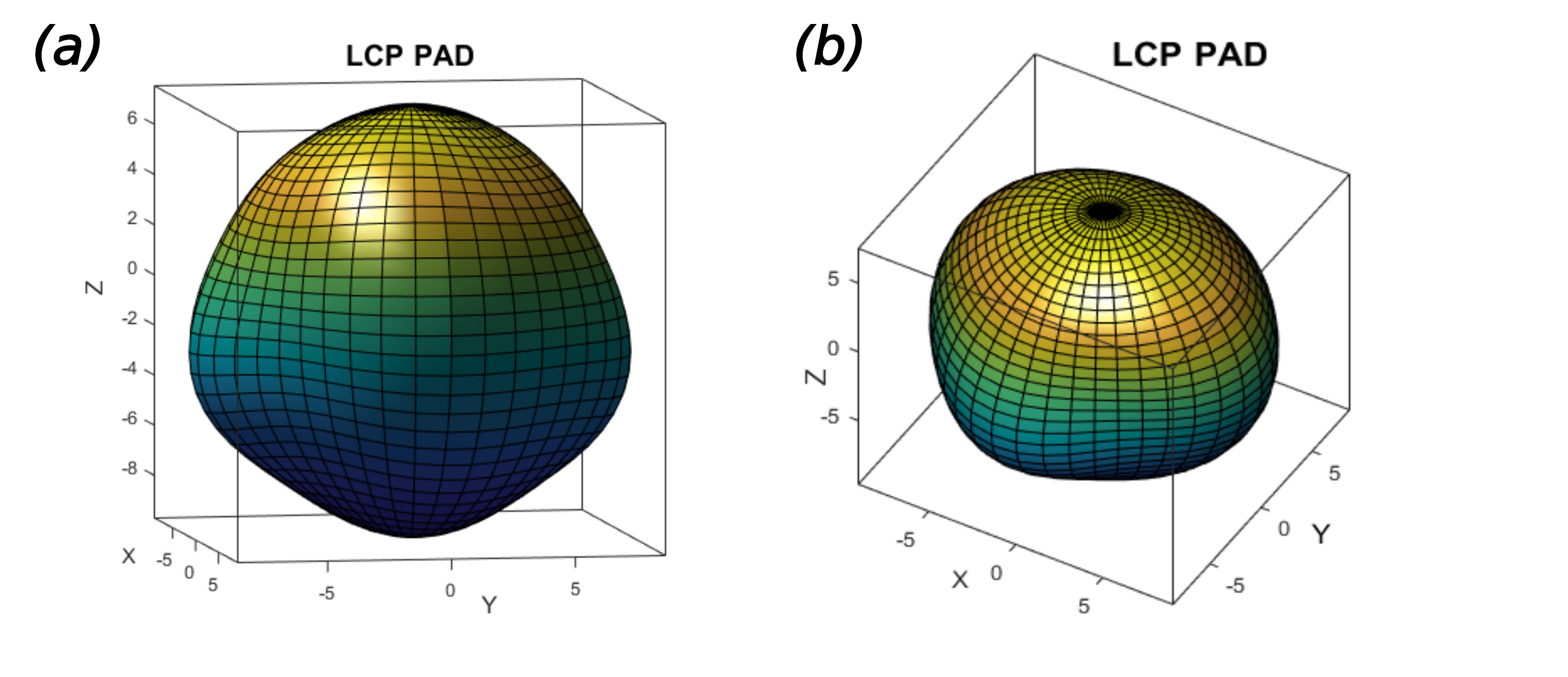}
	\caption{Calculated 3D PAD for the C$_1$ 1s ionization of the 3s Rydberg state of S-(+)-fenchone. The LCP polarized probe beam propagates along $Z$, the linear polarization of the pump beam lies along $Y$, and the detection axis along $X$.}
	\label{fig:lcpPAD}
\end{figure}

\begin{figure}
	\centering
	\includegraphics[width =0.7\textwidth]{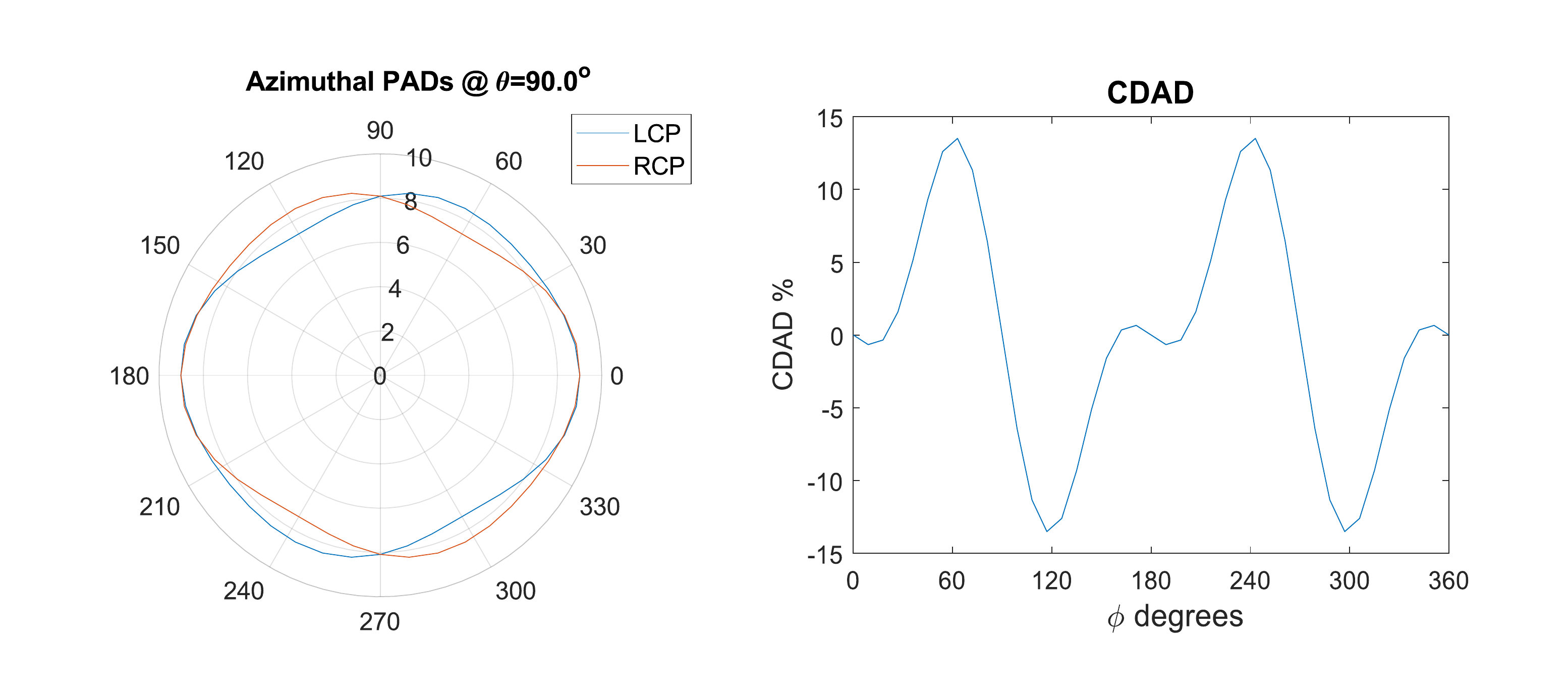}
	\caption{\label{fig:azimplt} Azimuthal angular distributions at $\theta = 90^o$ for the 3D PAD (Fig. \ref{fig:lcpPAD}) (a)  for LCP and RCP probe polarization . (b) the difference, or dichroism, LCP-RCP of these azimuthal distributions.}

\end{figure}

The full C$_1$ 1s$^{-1}$ 3D PAD calculation used to evaluate the consequences of alignment of the 3s Rydberg state by the linearly polarized two photon excitation beam is shown in Fig. \ref{fig:lcpPAD}. The forward-backward asymmetry along the CPL probe beam direction ($Z$) is clearly visible as is the loss of axial symmetry. This can be seen more clearly in the azimuthal plot Figure \ref{fig:azimplt}. It may be noted that the major axis of distortion lies neither perpendicular nor parallel to the linear polarization axis, $Y$, but rather at some intermediate angle. Equally noteworthy, the is a clear difference between distortion experienced with LCP or RCP \textit{probe} polarizations. At the polar angle $\theta_Z=90^o$ shown, PECD must necessarily be zero, having a $\cos\theta_Z$ dependence. Rather this is a form of dichroism (CDAD) where the experimental chirality arises from the mutually orthogonal arrangement of the CPL probe, axial alignment (linearly polarized pump), and detection axes, but \textit{not} directly from the molecular chirality.

\newpage

\end{document}